\documentclass[]{aastex631}
\usepackage{amsmath}
\usepackage{graphicx}
\usepackage{threeparttable}
\usepackage{txfonts}
\usepackage{soul}
\usepackage{bm} 
\usepackage{tabularx} 
\usepackage{booktabs}
\usepackage{varwidth}
\usepackage{graphicx}
\usepackage{CJKutf8}

\newcommand{\chinesename}{\textnormal{(}\begin{CJK*}{UTF8}{gbsn}张钊\end{CJK*}\textnormal{)}}

\shorttitle{Utilizing FRB DM to Probe the IGM Turbulence}
\shortauthors{Li et al.}

\begin{document}
\title{Utilizing Dispersion Measure of Fast Radio Bursts to Probe the Intergalactic Medium Turbulence}

\correspondingauthor{Fa-Yin Wang}
\email{fayinwang@nju.edu.cn}

\author[0009-0007-3326-7827]{Rui-Nan Li}
\affiliation{School of Astronomy and Space Science, Nanjing University Nanjing 210023, China}
\affiliation{Key Laboratory of Modern Astronomy and Astrophysics (Nanjing University) Ministry of Education, China}

\author[0000-0001-6869-2996]{Zhao Joseph Zhang \chinesename}
\affiliation{Theoretical Astrophysics, Department of Earth and Space Science, The University of Osaka, 1-1 Machikaneyama, Toyonaka, Osaka 560-0043, Japan}

\author[0000-0001-7457-8487]{Kentaro Nagamine}
\affiliation{Theoretical Astrophysics, Department of Earth and Space Science, The University of Osaka, 1-1 Machikaneyama, Toyonaka, Osaka 560-0043, Japan}
\affiliation{Theoretical Joint Research, Forefront Research Center, Graduate School of Science, The University of Osaka, Toyonaka, Osaka 560-0043, Japan}
\affiliation{Kavli IPMU (WPI), UTIAS, The University of Tokyo, Kashiwa, Chiba 277-8583, Japan}
\affiliation{Department of Physics \& Astronomy, University of Nevada, Las Vegas, 4505 S. Maryland Pkwy, Las Vegas, NV 89154-4002, USA}
\affiliation{Nevada Center for Astrophysics, University of Nevada, Las Vegas, 4505 S. Maryland Pkwy, Las Vegas, NV 89154-4002, USA}

\author[0000-0002-5712-6865]{Yuri Oku}
\affiliation{Center for Cosmology and Computational Astrophysics, the Institute for Advanced Study in Physics,  Zhejiang University, China} 

\author[0000-0001-6021-5933]{Qin Wu}
\affiliation{School of Astronomy and Space Science, Nanjing University Nanjing 210023, China}
\affiliation{Key Laboratory of Modern Astronomy and Astrophysics (Nanjing University) Ministry of Education, China}

\author[0000-0003-4157-7714]{Fa-Yin Wang}
\affiliation{School of Astronomy and Space Science, Nanjing University Nanjing 210023, China}
\affiliation{Key Laboratory of Modern Astronomy and Astrophysics (Nanjing University) Ministry of Education, China}

\begin{abstract}
Extragalactic fast radio bursts (FRBs) have emerged as powerful probes of turbulence within the intergalactic medium (IGM), a phenomenon that plays a crucial role in various cosmological and astrophysical processes. In this study, we employ the structure function (SF) analysis on the dispersion measures (DMs) of over 3,000 FRBs, leveraging the recently released CHIME/FRB Catalog 2 alongside previously observed sources. By comparing our results with mock datasets generated from cosmological simulations, we find excellent agreement at large angular separations. At small angular scales, our findings reveal a potential scaling behavior consistent with a two-dimensional (2D) Kolmogorov power spectrum. From this scaling, we constrain the turbulence outer scale to be on the order of several Mpc, which aligns with theoretical expectations, independent observations of the low-redshift IGM, and cosmological simulations. Ultimately, to conclusively confirm this Kolmogorov-like turbulent cascade and overcome current small-sample statistical limitations, a larger sample of FRBs with sub-arcsecond localization is required.
\end{abstract}


\section{Introduction} \label{sec:intro}
Turbulence plays a significant role across a wide range of astrophysical environments and spatial scales \citep{Armstrong1995, Biskamp2003, Schuecker2004, Elmegreen2004, Vogt2005, Chepurnov2010}. Its statistical properties can be quantified using a variety of tools, including power-spectrum analysis, correlation functions (CFs), and structure functions (SFs). In the interstellar medium (ISM) and the intracluster medium (ICM), turbulence has been constrained through multiple complementary observables \citep{Chepurnov2010, Lazarian2016, Xu2016, Xu2016_scattering, Xu2020b}. In contrast, turbulence in the intergalactic medium (IGM) remains poorly constrained because of the intrinsic faintness of the diffuse IGM and the difficulty of obtaining direct in-situ measurements.

From a theoretical perspective, IGM turbulence is expected to arise naturally from nonlinear structure formation and feedback-driven processes, generating vorticity, shocks, and non-thermal pressure support that can profoundly affect baryon dynamics on Mpc and sub-Mpc scales \citep{Ryu2008, Evoli2011, Rorai2017}. Observationally, clustering studies of both the highly ionized gas and the cooler HI medium reveal a spatial correlation length of a few Mpc \citep{Tejos2014, Finn2016}. Meanwhile, cosmological simulations provide a powerful avenue for characterizing such turbulent motions and their statistical signatures in regimes that are currently inaccessible to direct observations \citep{Zhu2010, Ursino2011, Iapichino2011, Zhu2011}. Nevertheless, direct observational constraints remain highly sought after to stringently test these theoretical predictions and to calibrate simulation-based inferences.

Fast radio bursts (FRBs) are millisecond-duration, high-brightness radio transients of unknown physical origin \citep{Lorimer2007, Xiao2021, Zhang2023Review}. As FRB signals propagate through ionized media, including the IGM, they experience frequency-dependent dispersion. The dispersion measure (DM), defined as the free-electron column density integrated along the line of sight (LOS), quantifies this effect. Observed FRB DMs range from a few tens to a few thousand $\rm{pc\,cm^{-3}}$ \citep{chimecatalog2}, far exceeding typical values for Galactic pulsars, indicating that most FRBs originate at extragalactic distances. For such sources, a substantial fraction of the measured DM is contributed by the IGM \citep{Ioka2003, McQuinn2014}, which can be used as cosmological probes \citep{ZhangZJ2021,Wu2022,YangKB2022,Gao2025}. Importantly, density fluctuations induced by IGM turbulence should imprint measurable spatial fluctuations in FRB DMs, enabling a statistical probe of IGM turbulence. 

\cite{Xu2020} performed a statistical study of IGM turbulence using SF analysis of the DMs of 122 FRBs, finding a tentative Kolmogorov-like scaling with angular separation at $\sim10^\circ$. However, the limited sample size led to large uncertainties. Subsequent work showed that such apparent scaling vanished after incorporating the first CHIME/FRB Catalog \citep{chime_cat}. The updated results suggested no power-law behavior at small angular separations, and highlighted a discrepancy of unknown origin between the directly measured SF and estimates derived from the CF \citep{Xu2021}. The empirically derived FRB SF was also found to be in good agreement with analytical approximations extracted from the IllustrisTNG300 cosmological simulations \citep{Nelson2018}. However, these findings await rigorous validation with larger FRB samples.

Recently, the CHIME/FRB Collaboration released the second CHIME/FRB Catalog \citep{chimecatalog2}, containing DMs for more than 3,000 unique FRB sources. In this work, we combine the second CHIME/FRB Catalog with previously published FRBs to construct an enlarged DM sample, and we apply SF analysis to investigate IGM turbulence. To interpret the observed SF with physically motivated models, we compare the FRB-derived SF with synthetic DM maps extracted from the {\small GADGET-4} CROCODILE simulations\footnote{\url{https://sites.google.com/view/crocodilesimulation/home}} \citep{Oku2024, ZhangZ2025}. The CROCODILE framework incorporates updated thermal AGN and revised supernova (SN) feedback models, providing a broad suite of runs with varied feedback parameters for systematic comparison. Because AGN-driven heating and outflows regulate baryon redistribution and the injection of turbulence in the intergalactic medium, this {\small GADGET-4}-based suite is particularly well-suited for robustly benchmarking the FRB DM structure function.


The structure of this paper is as follows. In Section \ref{sec:SF}, we briefly introduce the SF formalism for DM fluctuations and describe the FRB dataset and the simulation data. In Section \ref{sec:SF results}, we present the SF results and discuss their implications. Finally, we present discussions and summarize our conclusions in Section \ref{sec:dis}.

\section{Structure function and data set} \label{sec:SF}
The electron density fluctuations within the turbulent medium manifest as fluctuations in the DM, defined as $\delta \mathrm{DM}(\boldsymbol{X}) = \mathrm{DM}(\boldsymbol{X}) - \langle \mathrm{DM} \rangle$. Following \citet{Xu2021}, the CF of the FRB DM is modeled as:
\begin{equation}\label{equ:1}
\xi(R) =\left\langle\delta \mathrm{DM}\left(\boldsymbol{X}_{\mathbf{1}}\right) \delta \mathrm{DM}\left(\boldsymbol{X}_{\mathbf{2}}\right)\right\rangle,
\end{equation}
where $\boldsymbol{X}$ denotes the 2D position of the source on the sky plane. $R = |\boldsymbol{X_1}-\boldsymbol{X_2}|$ represents the projected transverse separation. For extragalactic sources such as FRBs, this transverse separation corresponds to an angular separation $\theta$ via $R = \theta L$, where $L$ denotes the effective distance from the observer to the source.
The SF of the FRB DM is defined as
\begin{equation}\label{equ:2}
\begin{aligned}
D(R) & =\left\langle\left[\mathrm{DM}\left(\boldsymbol{X}_{1}\right)-\mathrm{DM}\left(\boldsymbol{X}_{2}\right)\right]^{2}\right\rangle \\
& =\left\langle \mathrm{DM}(\boldsymbol{X}_{1})^2 + \mathrm{DM}(\boldsymbol{X}_{2})^2 - 2\mathrm{DM}(\boldsymbol{X}_{1}) \mathrm{DM}(\boldsymbol{X}_{2}) \right\rangle,
\end{aligned}
\end{equation}
which has the similar form of the SF of rotation measure \citep{Xu2016, Li2025a}. In the case of a stationary or homogeneous random process, the SF and CF convey equivalent information. Under these conditions, Equation \ref{equ:2} can be related to the CF via:
\begin{equation}\label{equ:3}
D(R) = C - 2 \xi(R),
\end{equation}
where $C = 2\langle \delta \mathrm{DM}^2 \rangle$ is a constant representing twice the variance of the DM fluctuations. However, it is important to note that in more general cases involving large-scale inhomogeneities, the SF and CF are not strictly interchangeable. The SF is generally preferred as it is less distorted by low-frequency trends (red noise) and offers higher accuracy for a given dataset size \citep{Schulz1981}. \citet{Xu2020} reported a discrepancy between the SF and CF derived from FRB data. We will investigate this discrepancy using a significantly larger sample in the following section.

We combine the previously reported unlocalized and localized FRB data from the Blinkverse FRB database \footnote{Blinkverse FRB database: \url{https://blinkverse.zero2x.org/}} \citep{blinkverse} and \cite{Li2025} with the second CHIME/FRB catalog \citep{chimecatalog2}. This combined catalog yields a total of 3738 unique FRB sources with measured DMs. The observed DM of an extragalactic FRB can be decomposed into four primary components:
\begin{equation}\label{equ:4}
\begin{aligned} 
\mathrm{DM}_{\text {obs}} = & \mathrm{DM}_{\mathrm{MW, ISM}} + \mathrm{DM}_{\mathrm{MW, halo}} + \mathrm{DM}_{\mathrm{IGM}} + \frac{\mathrm{DM}_{\text {host}} + \mathrm{DM}_{\text {source}}}{1+z},
\end{aligned}
\end{equation}
where $\mathrm{DM}_{\mathrm{MW, ISM}}$ represents the contribution from the Milky Way's interstellar medium, which can be estimated using Galactic electron density models such as NE2001 \citep{ne2001} or YMW16 \citep{ymw16}. The contribution from the Milky Way halo, $\mathrm{DM}_{\mathrm{MW, halo}}$, remains poorly constrained. The exact contributions from the host galaxy $\mathrm{DM}_{\text {host}}$ and local source environment $\mathrm{DM}_{\text {source}}$ are also difficult to disentangle without introducing significant bias. We therefore define $\rm{DM_{exc}}$ as: $\mathrm{DM}_{\mathrm{exc}} = \mathrm{DM}_{\mathrm{obs}} - \mathrm{DM}_{\mathrm{MW, ISM}}$.

To minimize potential contamination from the local circumburst environment, host galaxy contributions, or instrumental artifacts, we exclude FRBs with extreme dispersion measures ($\mathrm{DM} > 3000~\rm{pc~cm^{-3}}$) from our analysis. Additionally, FRB 200428 which is associated with the Galactic magnetar SGR 1935+2154 \citep{Bochenek2020, CHIME/FRB2020} is excluded to ensure a strictly extragalactic sample. 

\begin{figure} 
\centering
\includegraphics[width=180 mm]{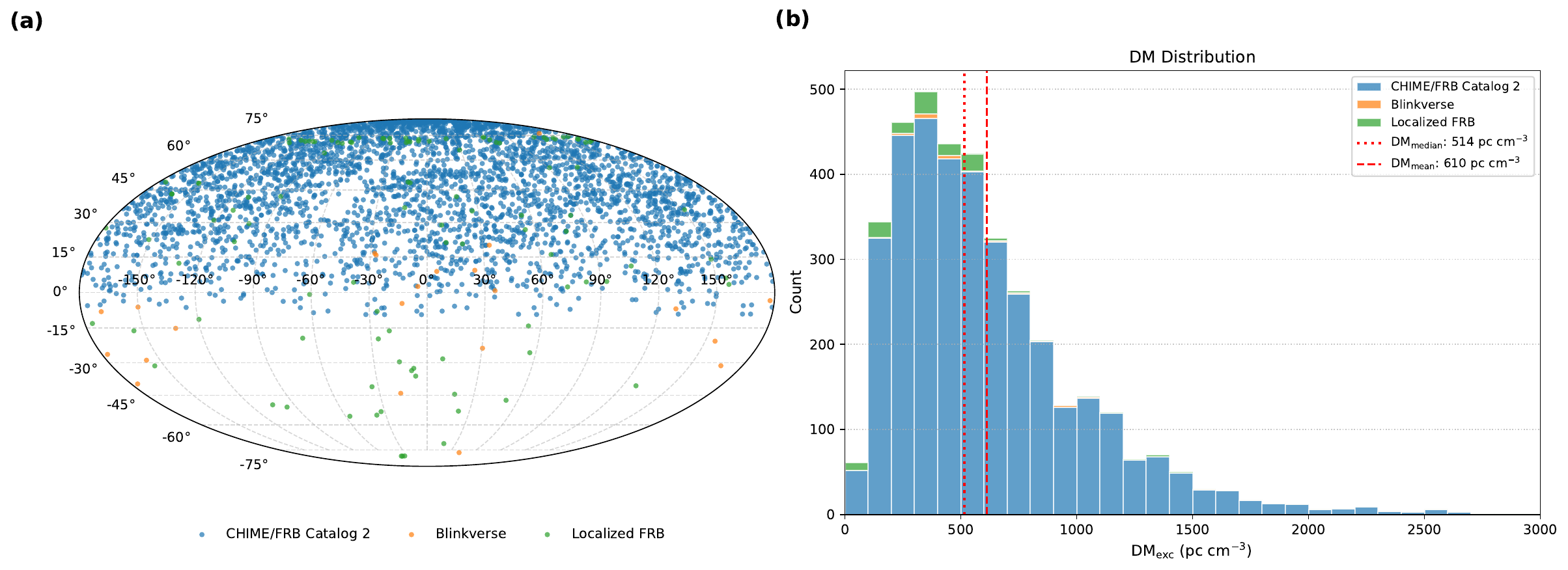}
\caption{Panel (a): Sky distribution of the FRB sample used in our analysis, after excluding sources with $\rm{DM_{exc}} > 3000~\rm{pc~cm^{-3}}$ and the Galactic source FRB 20200428A. FRBs appearing in multiple catalogs are represented by blue points. Panel (b): Histogram of the $\rm{DM_{exc}}$ distribution after extracting MW contribution based on NE2001 model for each FRB in this sample. The red dotted line denotes the median value and the dashed-dotted line represents the mean value. The dataset combines sources from the CHIME/FRB Catalog 2 \citep{chimecatalog2}, the Blinkverse database \citep{blinkverse}, and recent localized FRBs summarized in \cite{Li2025} and newly localized sources in \cite{Marazuela2026}. 
}
\label{Fig1}
\end{figure}

The sky position and $\rm{DM_{exc}}$ distributions of the final sample are presented in panels (a) and (b) of Figure \ref{Fig1}, respectively. This sample exhibits a strong spatial bias toward the Northern celestial hemisphere because the majority of events were detected by the CHIME telescope, which has a limited field of view (Dec $\leq-9.5^\circ$). Consequently, there is a paucity of sources in the Southern sky. However, according to the Cosmological Principle, which posits isotropy and homogeneity on Mega-parsec scales, we assume that the turbulent properties of the IGM derived from the Northern sky are representative of global statistics. Any hemispherical asymmetry in the IGM distribution is expected to be negligible compared to the stochastic density fluctuations accumulated over cosmological distances. We acknowledge that this non-uniform sky coverage could introduce systematics related to Galactic foreground subtraction. Although we subtracted the $\mathrm{DM}_{\mathrm{MW, ISM}}$ using standard electron density models (e.g., NE2001), residual model inaccuracies at Northern Galactic latitudes and $\mathrm{DM}_{\mathrm{MW, halo}}$ could theoretically propagate into our structure function analysis. Nevertheless, given that most sources are located at high Galactic latitudes and the path lengths are dominated by the IGM, these local residuals are likely sub-dominant. Future surveys covering the Southern sky, such as those by ASKAP \citep{ASKAP}, MeerKAT \citep{MeerKAT}, and the SKA \citep{SKA}, will be essential to cross-validate these findings and constrain any potential hemispherical systematics. 

To verify whether the non-uniform sample distribution significantly affects the SF analysis results, we also incorporate the cosmological simulation data into our analysis. Observations of several repeating FRBs have revealed temporal variations in their DMs \citep{Spitler2014, Spitler2018, Michilli2018, Wang2025, Niu2026}, likely driven by dynamically evolving circumburst environments. Because this localized DM contribution is variable and difficult to precisely decouple from the total observed DM, we adopt the time-averaged ${\rm DM_{exc}}$ for all repeating sources in our analysis.

As shown in panel (b) of Figure \ref{Fig1}, $\rm{DM_{exc}}$ of this sample of FRBs is centered at approximately $550~\rm{pc~cm^{-3}}$. Assuming a combined average contribution of $\sim150~\rm{pc~cm^{-3}}$ from the Milky Way halo, the host galaxy and local environment \citep{Zhang2020,Wang2022, Li2025}. The DM contributed by IGM is estimated to be around $400~\rm{pc~cm^{-3}}$. Based on the $\mathrm{{DM}}-z$ relation, where $\rm{{DM_{IGM}}} \sim (800-1000) \mathit{z}~\rm{pc~cm^{-3}}$, this implies that the majority of FRBs in the full sample are located at redshifts $z\lesssim0.5$ \citep{Deng2014, Zhang2018b, Macquart2020, Batten2021, ZhangZJ2021, Walker2024, Li2025}.
For the subsamples with DM upper-limit of $300~\rm{pc~cm^{-3}}$ and $400~\rm{pc~cm^{-3}}$, the corresponding redshift limits are below 0.2 and 0.3, respectively. However, the analytically expression for the CF modeled by \cite{Takahashi2021}, which was used for comparison in \cite{Xu2021} is strictly valid only for $z>0.3$ and $\theta\gtrsim1^\circ$. Consequently, to ensure robust results across the full redshift range of this larger samples, we employ cosmological simulations to perform a more precise comparison.

For the simulation analysis, we reconstructed grid-based gas-ion data from two $100~\mathrm{cMpc}/h$ CROCODILE boxes, \texttt{L100N1024\_Fiducial} and \texttt{L100N1024\_NoBH}. At the current stage, we use the completed $128^3\,(\mathrm{lv7})$ and $256^3\,(\mathrm{lv8})$ Cartesian meshes (i.e., $x_{\rm bin\_num}\times y_{\rm bin\_num}\times z_{\rm bin\_num}$), while the $512^3\,(\mathrm{lv9})$ runs are still in progress. As summarized in Table \ref{tab:crocodile_grid}, four runs with distinct parameters are evaluated. We confirm that these variations introduce no significant impact on the resulting SF. Ultimately, we select the \texttt{L100N1024\_Fiducial} grid at $256^3\,(\mathrm{lv8})$ with AGN feedback as our fiducial model for comparison with the FRB data, which corresponds to a spatial resolution of $\simeq 0.39~\mathrm{cMpc}/h$ per cell. All simulation snapshots utilized herein cover the redshift range from $z = 0$ to $2$. The DM is integrated along each simulated LOS up to a specific distance determined by the redshift resolution. This approach allows us to artificially place mock FRBs at any arbitrary location within the redshift range of $z = 0 - 2$ and directly obtain their corresponding DMs.

\begin{figure}
\centering
\includegraphics[width=0.98\textwidth]{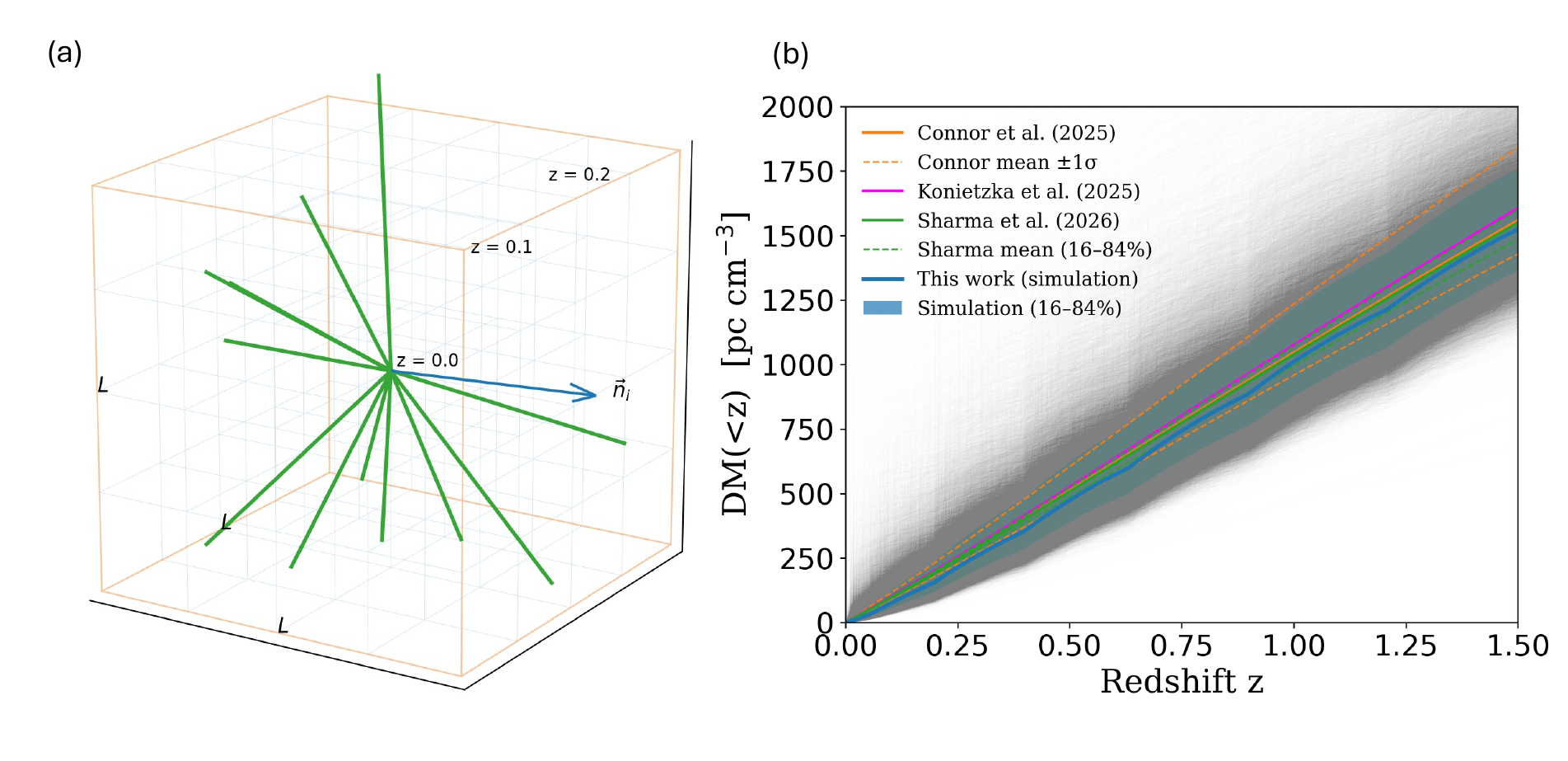}
\caption{
Panel (a): Schematic illustration of the light-cone DM construction in periodic CROCODILE simulation boxes. Green rays denote representative lines of sight $\hat{n}_i$ launched from the observer at $z=0$, and outer boxes indicate stacking toward higher redshift. Panel (b): Comparison of the diffuse DM evolution, derived from 50{,}000 uniformly sampled sightline directions.  The blue solid curve and shaded region indicate the simulation median and the 16th--84th percentile sightline scatter, respectively.  The orange, magenta, and green curves show the corresponding reference relations from \citet{Connor2024}, \citet{Konietzka2025}, and \citet{Sharma2026} The data are extracted from the \texttt{L100N1024\_Fiducial} simulation grid at $256^3\,(\mathrm{lv8})$ incorporating AGN feedback.
}
\label{Fig2}
\end{figure}

\paragraph{Light-cone construction from grid data.}
All subsequent calculations are performed on the reconstructed grid data. We construct FRB DM light-cones by stacking periodic simulation boxes along fixed LOS, without rotating the grid. As illustrated in panel (a) of Figure \ref{Fig2}, an observer is placed at a fixed position inside the full simulation volume. We select this observer in a relatively diffuse region with local gas density in the range $10^{-4}$--$10^{-3}~\mathrm{g\,cm^{-3}}$, so that strong local contamination analogous to Milky Way ISM contributions is minimized in the cosmological DM analysis.

From this observer, we uniformly generate 50,000 sightline directions $\hat{n}_i$ to sample the full sky. 
For each direction, the ray propagates radially and intersects successive simulation boxes.  
The physical trajectory is mapped into the periodic simulation volume through coordinate wrapping, such that
\begin{equation}
\mathbf{x}_{\rm sim} = \mathbf{x}_{\rm world} \bmod L_{\rm box},
\end{equation}
where $\mathbf{x}_{\rm world} = \mathbf{x}_{\rm obs} + \chi(z)\,\hat{n}_i$ denotes the physical comoving position along the LOS in the light-cone geometry, and $\mathbf{x}_{\rm sim}$ is the corresponding position mapped into the finite simulation volume of size $L_{\rm box}$. Since the rays propagate obliquely through the periodic volume rather than re-entering the box along the same axis, re-sampling the same cell requires the wrapped trajectory to return within one-cell tolerance in all three coordinates simultaneously. This mapping allows each ray to continuously sample the large-scale structure while preserving the intrinsic density field of each snapshot and strongly suppressing same-cell recurrence, as detailed in Appendix~\ref{app:recurrence}.

Snapshots at different redshifts are stacked sequentially along the same $\hat{n}_i$: nearby segments are taken from lower-redshift outputs, while more distant segments are taken from higher-redshift outputs. 
The corresponding DM evolution with redshift for these 50,000 sightlines is shown in panel (b) of Figure~\ref{Fig2}.

To suppress artificial discontinuities at snapshot interfaces, we introduce a narrow overlap region between adjacent boxes and smoothly blend the electron density across the boundary before integrating DM. 
This approach preserves spatial coherence along each ray and avoids non-physical discontinuities induced by transitions between neighboring snapshots, enabling the direct construction of cumulative DM profiles and full-sky DM maps as a function of redshift. The full analysis pipeline, including light-cone construction and line-of-sight sampling, is implemented in our \textit{FALCON} module within the publicly available \textit{ARCOS} codebase\footnote{\url{https://github.com/zhaojosephzhang/ARCOS}}.

\begin{table*}[t]
\centering
\begin{threeparttable}
\caption{Summary of CROCODILE simulation boxes and currently available reconstructed grids.}
\label{tab:crocodile_grid}
\begin{tabular}{lcccc}
\toprule
Run & Box size & Grid size & Cell size & Used for comparison \\
 & ($\mathrm{cMpc}/h$) & ($x_{\rm bin\_num}\times y_{\rm bin\_num}\times z_{\rm bin\_num}$) & ($\mathrm{cMpc}/h$) & \\
\midrule
\texttt{L100N1024\_Fiducial} & 100 & $128^3\,(\mathrm{lv7})$ & $100/128\approx0.78$ &  \\
\texttt{L100N1024\_Fiducial} & 100 & $256^3\,(\mathrm{lv8})$ & $100/256\approx0.39$ & $\checkmark$ \\
\texttt{L100N1024\_NoBH} & 100 & $128^3\,(\mathrm{lv7})$ & $100/128\approx0.78$ &  \\
\texttt{L100N1024\_NoBH} & 100 & $256^3\,(\mathrm{lv8})$ & $100/256\approx0.39$ &  \\
\bottomrule
\end{tabular}
\begin{tablenotes}
\item Note. $\mathrm{lv}N$ denotes a grid level with $2^N$ cells per axis; therefore $\mathrm{lv7}=128$ and $\mathrm{lv8}=256$.
\end{tablenotes}
\end{threeparttable}
\end{table*}


\begin{figure} 
\centering
\includegraphics[width=180 mm]{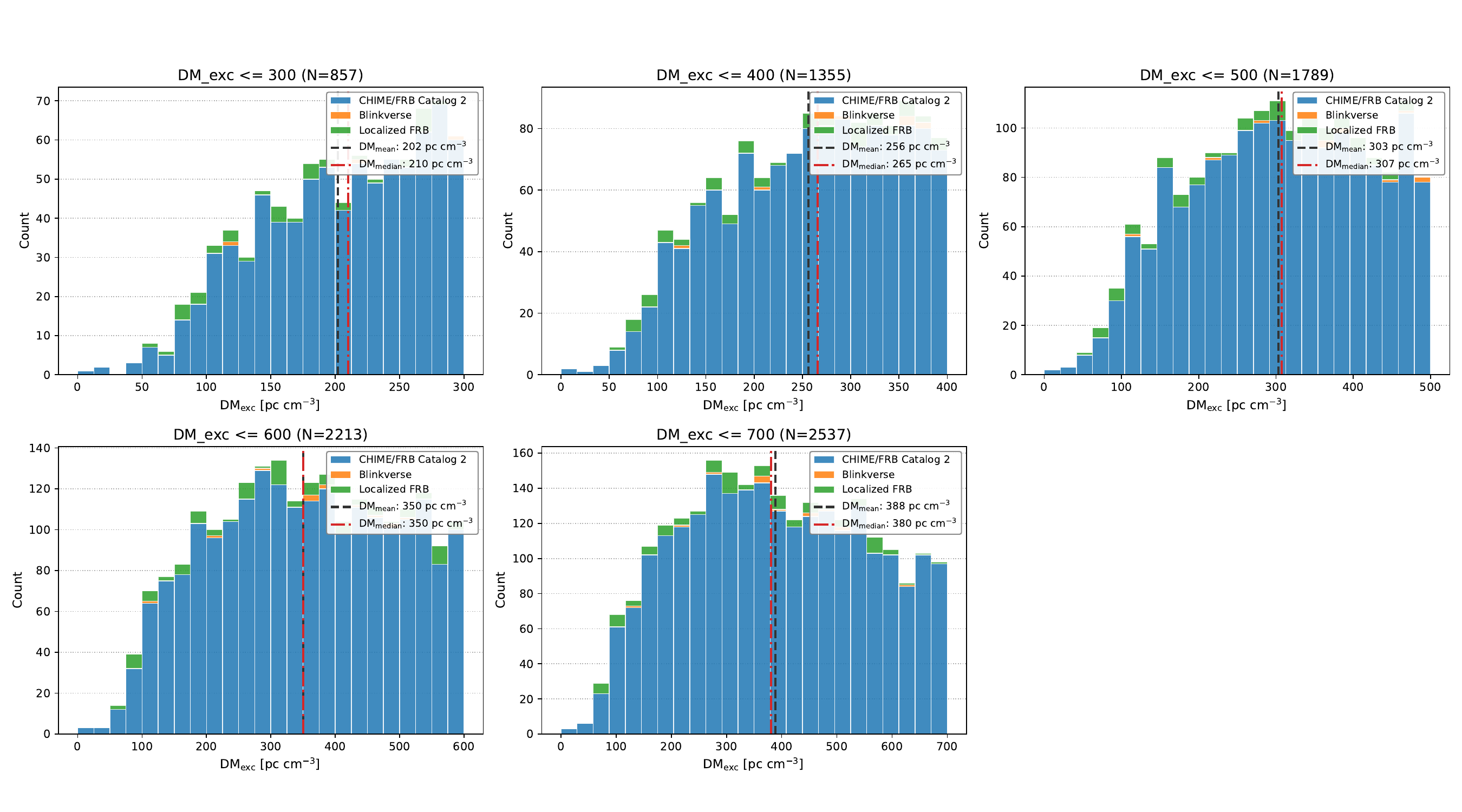}
\caption{Distributions of $\rm DM_{exc}$ for the composite FRB sample, partitioned by varying upper-limit threshold cuts. Sub-panels illustrate the stacked histograms of the sample truncated at $\rm DM_{exc} \le 300$, $400$, $500$, $600$, and $700~{\rm pc~cm^{-3}}$, with the corresponding total number of retained sources ($N$) indicated. The heterogeneous origins of the sample are delineated by color: the CHIME/FRB Catalog 2 (blue), the Blinkverse database (orange), and the precisely localized FRB sample (green). The statistical mean ($\rm DM_{mean}$) and median ($\rm DM_{median}$) of each subsample are denoted by the vertical black dashed and red dash-dotted lines, respectively.
}
\label{Fig3}
\end{figure}

\begin{figure} 
\centering
\includegraphics[width=180 mm]{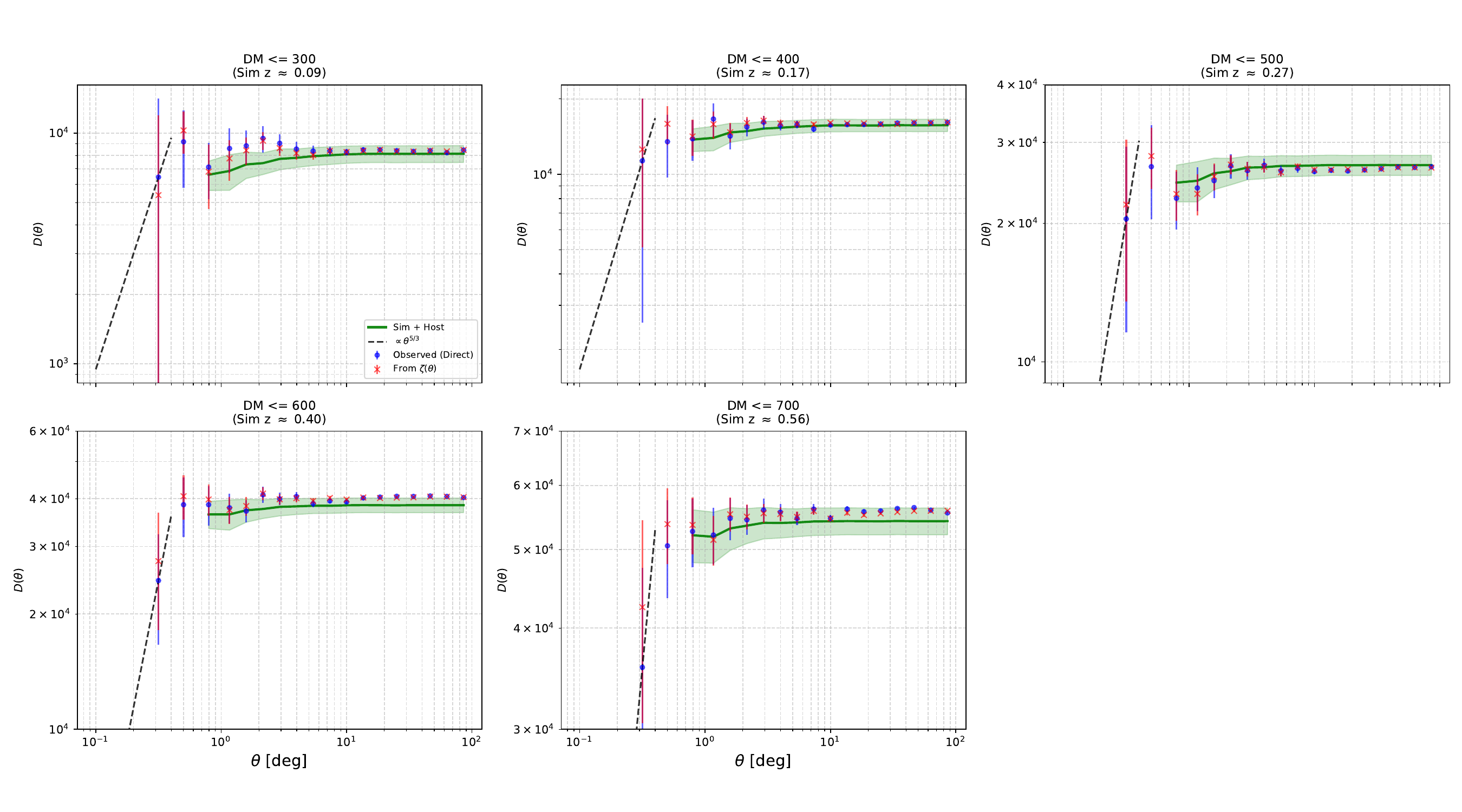}
\caption{The $\mathrm{DM_{exc}}$ structure function results for the FRB sample across various DM upper limits. The blue circles and red crosses represent the SF calculated directly from FRB pairs and derived from the correlation function, respectively, with error bars denoting $95\%$ confidence intervals. Source pairs with angular separations smaller than their localization errors are excluded to prevent spurious small-scale signals. The solid green lines and shaded $1\sigma$ regions represent the simulated SFs, which are truncated at redshifts selected to match the observed SF amplitude at $\theta \gtrsim 1^\circ$. The black dashed lines indicate the theoretical $5/3$ scaling of Kolmogorov turbulence, fitted to the data at $0.05^\circ < \theta < 0.5^\circ$.
}
\label{Fig4}
\end{figure}

\section{Structure function results of FRB DM} \label{sec:SF results}
In this analysis, we assume that the turbulent medium is distributed along the entire LOS, as is characteristic of the IGM. Under this framework, $L$ represents the total length of the turbulent volume encompassing both the sources and the observer. For extragalactic sources such as FRBs, the effective LOS distances span the range $[L_0, L]$, where $L_0$ denotes the distance to the nearest FRB in our sample. The SF can thus be analytically expressed as a function of the transverse distance $R$ \citep{Xu2020}:

\begin{equation}\label{equ:5}
\begin{aligned}
D(R)\approx \begin{cases}2 \langle \delta n_e^2 \rangle L_i^{-m} (L + L_0) R^{m+1} + \frac{1}{3} \langle \delta n_e^2 \rangle (L - L_0)^2, & R < L_i \\
2 \langle \delta n_e^2 \rangle L_i (L + L_0) + \frac{1}{3} \langle \delta n_e^2 \rangle (L - L_0)^2,  & R > L_i,\end{cases}
\end{aligned}
\end{equation}
where $\langle \delta n_e^2 \rangle$ denotes the variance of the electron density fluctuations, and $L_i$ represents the turbulence correlation length. For two-dimensional (2D) Kolmogorov turbulence, the power-law index corresponds to $m+1 = 5/3$.
According to Equation (\ref{equ:5}), variations in the source distances within the DM sample introduce an additional variance term to the SF \citep{Xu2020}. To mitigate this statistical bias, following \cite{Xu2021}, we impose a series of upper-limit DM thresholds (e.g., $300, 400, 500, 600$, and $700~{\rm pc~cm^{-3}}$) and derive the SF independently for each subsample. Naturally, a lower DM threshold yields a smaller SF amplitude due to the reduced integration path length. The ${\rm DM_{exc}}$ distributions for these subsamples are presented in Figure \ref{Fig3}. Furthermore, to prevent spurious correlations at small scales caused by the large positional uncertainties of non-localized FRBs, we strictly exclude any source pairs with angular separations $\theta$ smaller than their respective localization errors. 

The resulting SFs for the different DM thresholds are presented in Figure \ref{Fig4}, with error bars representing the $95\%$ confidence intervals for each angular bin. To account for small-sample effects, we employed a Student’s t-distribution with $n-1$ degrees of freedom for bins containing fewer than 30 pairs, while defaulting to a normal distribution for larger samples. This approach effectively mitigates the systematic underestimation of statistical uncertainties inherent to small sample sizes. The black dashed lines indicate the theoretical $5/3$ scaling expected for 2D Kolmogorov turbulence. The blue and red data points denote the SFs calculated directly from the data and those derived from the CF via Equation (\ref{equ:3}), respectively. The green solid lines represent the SFs derived from cosmological simulations. We align the simulations with the observational data by visually matching their amplitudes at $\theta \gtrsim 1^\circ$. Specifically, we truncate the simulation at an redshift that reproduces this observed amplitude. To account for cosmic variance, the simulated SFs are processed using a bootstrap resampling method. By randomly drawing 8,000 LOS over 1,000 iterations, we computed both the expected mean SF and its associated $1\sigma$ uncertainty envelope which quantifies the theoretical sample variance (i.e., cosmic variance) expected for our FRB dataset. Assuming statistical independence among the 8,000 randomly selected LOS may slightly underestimate the uncertainties due to spatial correlations. We leave the mitigation of this effect to future work, which will rely on spatial resampling methods (e.g., Jackknife) or the derivation of the full cosmological covariance matrix.

As shown in Figure \ref{Fig4}, the observational results exhibit good consistency with the cosmological simulations at $\theta \gtrsim 1^\circ$. At these scales, both observed and simulated SFs reach a plateau. This aligns with theoretical expectations that once the physical separation exceeds the turbulence outer scale, DMs become uncorrelated, causing the SF to flatten \citep{Lazarian2016, Xu2020, Xu2021}. This agreement implies that any spatial anisotropy within our FRB sample has a negligible impact on the global SF analysis. However, the currently available grid data do not provide sufficient angular resolution to robustly probe the SF at smaller scales ($\theta \lesssim 0.7^\circ$). Consequently, direct comparisons at the smallest separations remain inconclusive, and we defer a more detailed grid-refined treatment to future work.

When angular separations approach $0.4^\circ$, all subsamples exhibit a tentative decline in the SF, although the trend is subject to larger statistical uncertainties owing to the reduced number of pair counts per angular bin. The slope of this decline is consistent with the $5/3$ scaling predicted for 2D Kolmogorov turbulence, with an apparent break scale occurring at roughly $\theta_{\rm break} \approx 0.4^\circ$. Equation (\ref{equ:5}) indicates that the physical scale at which the SF flattens corresponds directly to the turbulence correlation length $L_i=\theta_{\rm {break}} L$. 
We first estimate the effective redshift $\langle z \rangle$ for each subsample based on its mean excess dispersion measure, $\langle {\rm DM_{exc}} \rangle$. According to Equation (\ref{equ:4}), we model the dispersion measure budget as $\langle {\rm DM_{exc}} \rangle = \langle {\rm DM_{IGM}}(z) \rangle + {\rm DM_{MW,halo}} + \langle {\rm DM_{local}} \rangle /(1+z)$. To account for the unmodeled foreground and local components, we adopt a constant Milky Way halo contribution of ${\rm DM_{MW,halo}} \approx 40~{\rm pc~cm^{-3}}$ \citep{Prochaska2019} and assume a representative rest-frame host galaxy contribution of $\langle {\rm DM_{local}} \rangle \approx 50~{\rm pc~cm^{-3}}$ \citep{Macquart2020,Zhang2020}. While individual FRB host environments exhibit significant DM variations, employing this mean-value approximation is robust for estimating the ensemble-average redshifts of our adequately sized subsamples. By adopting the empirical Macquart relation \citep{Macquart2020} for the IGM contribution ($\langle {\rm DM_{IGM}} \rangle \approx 855z~{\rm pc~cm^{-3}}$), we numerically solve the equation to obtain the effective redshift $\langle z \rangle$ for each DM threshold cut. Finally, utilizing the angular diameter distance $D_A(\langle z \rangle)$ derived from the standard Planck cosmology \citep{Planck2020}, the angular break scale translates to a physical transverse correlation length spanning $L_i \approx 3.8 - 7.4~{\rm Mpc}$ across our subsamples. These refined values are in agreement with theoretical expectations, independent estimates and cosmological simulation results for the outer scale of turbulence in the IGM \citep{Ryu2008, Zhu2010, Iapichino2011, Zhu2011, Ursino2011, Tejos2014, Finn2016, Evoli2011, Rorai2017}. Adopting alternative values for these other constants has a negligible impact on our final conclusions which is a advantage of SF.

\section{Discussion and summary} \label{sec:dis}
A discrepancy between the directly measured SF and the SF derived from the CF at angular scales $\theta < 10^\circ$ was previously reported by \cite{Xu2021}. This deviation was attributed to either statistical uncertainties or genuine density inhomogeneities on scales of $\sim 100~{\rm Mpc}$. However, such a discrepancy is not observed in our analysis, which incorporates a substantially larger FRB sample. This suggests that the previously noted inconsistency was primarily a statistical artifact driven by the limited number of FRB pairs at small angular separations. As shown across all panels in Figure \ref{Fig4}, while minor deviations between the two estimators occasionally appear in bins with the smallest sample sizes, but they remain in good agreement within the estimated error bars.

At small angular scales ($\theta \lesssim 0.5^\circ$), the sparse number of FRB pairs leads to substantial statistical uncertainties. While our conservative $95\%$ confidence intervals do not formally exclude a flat SF, the central estimates (i.e., mean value) systematically exhibit a distinct rising trend rather than random noise fluctuations. As the maximum likelihood estimators for their respective bins, these central values objectively trace the most probable underlying scaling. Strikingly, this rising slope is consistent with the $5/3$ power-law expected from 2D Kolmogorov turbulence. Because this $5/3$-like signature persistently appears across multiple independent DM subsamples, we suggest it may reflect a genuine physical origin probing the inertial range of the IGM turbulence cascade rather than mere statistical coincidence. Future accumulations of FRBs with sub-arcsecond localizations are essential to suppress these small-scale statistical errors and robustly constrain the IGM turbulence spectrum.

As shown in Figure \ref{Fig4}, we note that the exact location of the break scale, $\theta_{\rm break}$, cannot be resolved with infinite precision due to our discrete and extremely limited angular binning at $\theta \lesssim 0.5^\circ$. Because the decaying trend is primarily captured by a single innermost bin, the true turnover scale should intrinsically lie somewhere between the central value of this declining bin and that of its adjacent plateaued bin. Consequently, $\theta_{\rm break}$ is more accurately represented by an angular interval rather than a single discrete value. Adopting a conservative range of $\theta_{\rm break} \in [0.2^\circ, 0.5^\circ]$, the inferred values of $L_i$ can be interpreted as a reasonable bounding range of $L_i \approx 1.9 - 9.2~{\rm Mpc}$ for the IGM turbulence outer scale. 

At present, our simulation-based inference is limited by the resolution of the currently available reconstructed grids, primarily the $256^3$ ($\mathrm{lv8}$) products used in this work. In future work, higher-resolution grids (e.g., $\mathrm{lv9}$ and $\mathrm{lv10}$) will enable us to probe the SF behavior at smaller angular separations with improved robustness. We also find that the current differences between the \texttt{Fiducial} and \texttt{NoBH} cases are modest. A plausible explanation is that the present AGN prescription mainly includes thermal feedback; once AGN kinetic (jet-like) feedback is incorporated, a more pronounced divergence between these scenarios may emerge. 

In the present analysis we do not explicitly model the contribution from intervening halos. This is primarily because current observational data preclude the reliable identification and subtraction of all foreground halos along individual lines of sight. Consequently, halo gas intercepted by certain sightlines may introduce additional DM fluctuations. We expect these contributions to act primarily as an uncorrelated stochastic component in the SF, effectively adding marginal noise to the inferred turbulence signal.

In addition, our high-redshift simulation products are still being generated. Although this paper focuses on the redshift range most comparable to the current FRB observations (mainly $z\lesssim1$), future upgrades in radio and optical facilities are expected to substantially increase the number of detected high-redshift FRBs. The extended high-$z$ simulation datasets will therefore provide predictive $D(\theta)$ trends for that regime and offer a direct basis for future observational tests.

\begin{figure} 
\centering
\includegraphics[width=180 mm]{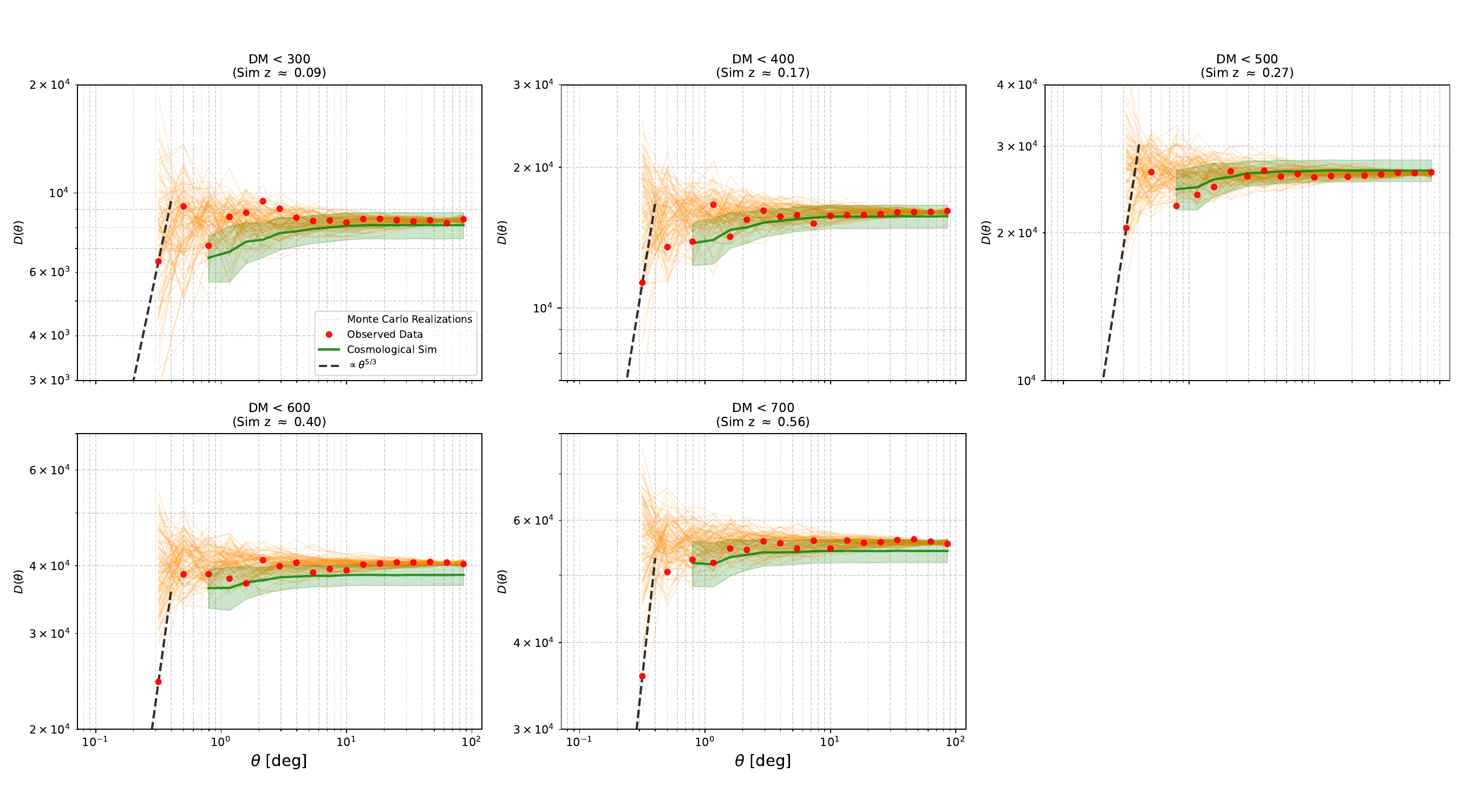}
\caption{Same as Figure \ref{Fig4}. 100 Monte Carlo permutation tests were performed by randomly shuffling the DM values while preserving the original FRB sky coordinates. The resulting null-hypothesis realizations are shown as thin orange lines.
}
\label{Fig5}
\end{figure}

To evaluate the statistical significance of the Kolmogorov-like scaling ($\propto \theta^{5/3}$) observed at small angular separations ($\theta \lesssim 0.5^\circ$), we performed a Monte Carlo permutation test following a methodology similar to that of \cite{Xu2021}. In this test, we generated 100 mock realizations for each DM upper-limit subsample. For each realization, we randomly shuffled the ${\rm DM_{exc}}$ values among the FRBs while preserving their true sky coordinates. This procedure effectively destroys any intrinsic spatial correlation of the IGM density field, yet maintains the identical geometrical distribution, angular pair-count statistics, and DM variance of the observed sample. The results are shown in Figure \ref{Fig5}. At $\theta \lesssim 0.5^\circ$, the observed SFs for subsamples with DM upper limits of $300$, $400$, and $500~{\rm pc~cm^{-3}}$ show no clear evidence of deviating from null-hypothesis envelope likely owing to the small pair counts. However, for subsamples with DM thresholds $> 500~{\rm pc~cm^{-3}}$, the observed SF significantly deviates from the envelope of the randomized mock data. Because these higher-DM subsamples contain a larger number of FRB pairs in their innermost angular bins, their statistical robustness should be correspondingly higher. Consequently, the distinct deviation from the null-hypothesis envelope supports the conclusion that the observed small-scale correlation is a genuine physical signature of IGM turbulence, rather than a random statistical fluctuation.

In this study, we incorporate data from the CHIME/FRB Catalog 2 alongside previously released FRB samples to investigate density fluctuations associated with IGM turbulence using SF analysis. To mitigate the statistical biases introduced by the broad distribution of source distances within the sample, we independently apply the SF analysis to multiple subsamples defined by varying upper-limit DM thresholds. Across all subsamples, we observe a systematic, monotonically increasing trend in the SF at small angular separations ($\theta \lesssim 0.5^\circ$). The slopes of these trends are highly consistent with the $5/3$ power-law expected for 2D Kolmogorov turbulence. Based on the mean and median ${\rm DM_{exc}}$ values of each subsample, we constrain the turbulence correlation length of the IGM to be $L_i \approx 3.8 - 7.4~{\rm Mpc}$. This inferred scale is in agreement with independent estimates derived via alternative methods in the literature. Furthermore, we apply the ARCO analysis code to the CROCODILE cosmological simulation to generate to generate cosmological simulations of FRB DMs. The SFs derived from the simulated datasets exhibit strong consistency with our observational results. However, the limited number of FRB pairs at angular separations below $0.5^\circ$ currently results in substantial statistical uncertainties. Besides, the current sample size of precisely localized FRBs remains insufficient for a definitive measurement. Future accumulations of FRBs with robust localizations are essential to strictly test the observed Kolmogorov scaling and further constrain the properties of IGM turbulence.

\section*{Acknowledgments}
This work was supported by the National Natural Science Foundation of China (grant Nos. 12494575, 12273009, 12447115 and 12503050).
Qin Wu is supported by the China Postdoctoral Science Foundation (CPSF) (grant Nos. GZB20240308, 2025T180875 and 2025M773199). 

\appendix
\section{Statistical Properties of Periodic Light-Cone Sampling}
\label{app:recurrence}

In this Appendix, we examine the statistical properties of line-of-sight sampling in a periodic simulation volume, and clarify the implications for structure repetition and the choice of light-cone construction method.

\subsection{Exact Recurrence in Continuous Space}

For a periodic box of size $L_{\rm box}$, an exact recurrence to the same spatial position requires that the displacement along the LoS satisfies
\begin{equation}
s \hat{n} = (m_x, m_y, m_z)\, L_{\rm box},
\end{equation}
where $s$ is the comoving distance along the line of sight, where $(m_x,m_y,m_z)$ are integers. This implies that the direction vector must satisfy
\begin{equation}
\hat{n} = \frac{1}{N}(m_x,m_y,m_z), \quad m_x^2+m_y^2+m_z^2=N^2.
\end{equation}

Such directions form a countable subset on the unit sphere. Therefore, under a continuous full-sky sampling, the probability of exact recurrence is strictly zero.

\subsection{Recurrence Scale in a Discretized Periodic Volume}
\label{app:recurrence_scale}

In our light-cone construction, each LoS propagates continuously through a periodic simulation volume along a fixed direction $\hat{n}=(n_x,n_y,n_z)$. 
Therefore, recurrence is not a random-sampling problem, but a geometric condition imposed on a correlated one-dimensional trajectory embedded in a three-dimensional periodic box.

For a box of size $L_{\rm box}$ and grid cell size $\Delta x=L_{\rm box}/N_{\rm grid}$, we define the same-cell recurrence scale as
\begin{equation}
L_{\rm recur}^{\rm cell}
=
\inf\left\{
s>0:\exists \mathbf{m}\in\mathbb{Z}^3,\ 
\max_i |s n_i - m_i L_{\rm box}| < \Delta x
\right\},
\end{equation}
where $\mathbf{m}=(m_x,m_y,m_z)$ is an integer triplet. 
This condition requires the periodic displacement of the trajectory to re-align within one-cell tolerance simultaneously in all three spatial dimensions.

For our simulation setup, $L_{\rm box}=100\,{\rm cMpc}$ and $N_{\rm grid}=256$, giving
\begin{equation}
\Delta x = \frac{L_{\rm box}}{N_{\rm grid}} \simeq 0.39\,{\rm cMpc}.
\end{equation}

For generic oblique sightline directions, the requirement of three-dimensional simultaneous alignment strongly delays recurrence. 
A characteristic scale can be estimated as
\begin{equation}
L_{\rm recur}^{\rm cell}
\sim
L_{\rm box}\left(\frac{L_{\rm box}}{\Delta x}\right)^2
=
L_{\rm box} N_{\rm grid}^2
\sim
100\times 256^2\,{\rm Mpc}
\simeq 6.6\times10^6\,{\rm cMpc}.
\end{equation}

For comparison, the comoving distance to $z=1$ is $\chi(z=1)\simeq 3.4\,{\rm cGpc}$, which is orders of magnitude smaller than the characteristic recurrence scale above. 
A direct Monte Carlo estimate based on random full-sky sightline directions yields a recurrence probability of $p_{\rm recur}\sim 1.84\times10^{-3}$ per sightline within this distance. 
For a sample of 50,000 sightlines, this corresponds to $\sim 10^2$ effective recurrence events.

Therefore, same-cell recurrence is not strictly absent, but remains rare within the redshift range relevant to our FRB sample ($z\lesssim1$), and is not expected to significantly affect the statistical properties of the measured DM fluctuations.

We note that re-entering the same cell does not imply that the subsequent trajectory must be strictly identical. 
Because the ray propagation is continuous, different visits to the same cell may have different entry and exit points. 
However, since the propagation direction $\hat{n}$ remains fixed, the local sampling geometry is nearly unchanged, and the corresponding density structures and large-scale information carried by the trajectory remain highly correlated. 
For this reason, we regard such events as effective recurrences for the purpose of assessing potential repetition bias.

\subsection{Implications for Light-Cone Construction}

Two main strategies are commonly adopted to mitigate artificial repetition in periodic simulations:

(i) Non-rotating continuous mapping:  
In this approach, each sightline propagates continuously through the periodic volume without random reorientation. This preserves spatial coherence along the LoS, ensuring that the sampled density field remains physically continuous.

(ii) Rotating-box approach:  
In this approach, each radial shell is randomly rotated and translated. This suppresses repetition of large-scale structures more efficiently and effectively increases the diversity of sampled configurations by reusing the simulation box under different orientations. In this sense, the rotating-box method can be viewed as generating multiple pseudo-independent realizations from a single simulation volume.

However, these advantages come at a cost. Random reorientation between shells introduces geometric discontinuities at shell boundaries, as the underlying density field is no longer continuous. These discontinuities cannot be fully removed by smoothing techniques and may introduce non-physical fluctuations along the LoS, potentially contaminating statistical measurements such as the DM structure function.

By contrast, the non-rotating method avoids such discontinuities and is therefore better suited for studies that rely on coherent line-of-sight statistics, such as probing IGM turbulence through DM fluctuations.

We therefore adopt the non-rotating approach as a conservative choice, prioritizing physical consistency over maximizing statistical reuse of the simulation volume.

\bibliographystyle{aasjournal}
\bibliography{ref1}

@ARTICLE{Wang2025,
       author = {{Wang}, F.~Y. and {Lan}, H.~T. and {Zhao}, Z.~Y. and {Wu}, Q. and {Feng}, Y. and {Yi}, S.~X. and {Dai}, Z.~G. and {Cheng}, K.~S.},
        title = "{Evidence of young magnetars in massive binary embedded in a supernova remnant as sources of active fast radio bursts}",
      journal = {arXiv e-prints},
         year = 2025,
        month = dec,
          eid = {arXiv:2512.07140},
        pages = {arXiv:2512.07140},
archivePrefix = {arXiv},
       eprint = {2512.07140},
 primaryClass = {astro-ph.HE},
       adsurl = {https://ui.adsabs.harvard.edu/abs/2025arXiv251207140W}
}

@ARTICLE{Gao2025,
       author = {{Gao}, D.~H. and {Wu}, Q. and {Hu}, J.~P. and {Yi}, S.~X. and {Zhou}, X. and {Wang}, F.~Y. and {Dai}, Z.~G.},
        title = "{Measuring the Hubble constant using localized and nonlocalized fast radio bursts}",
      journal = {\aap},
         year = 2025,
        month = jun,
       volume = {698},
          eid = {A215},
        pages = {A215},
          doi = {10.1051/0004-6361/202453006},
archivePrefix = {arXiv},
       eprint = {2410.03994},
 primaryClass = {astro-ph.CO},
       adsurl = {https://ui.adsabs.harvard.edu/abs/2025A&A...698A.215G}
}

@ARTICLE{Li2025a,
       author = {{Li}, Rui-Nan and {Zhao}, Zhen-Yin and {Wu}, Qin and {Yi}, Shuang-Xi and {Wang}, Fa-Yin},
        title = "{Structure Functions of Rotation Measures Revealing the Origin of Fast Radio Bursts}",
      journal = {\apjl},
         year = 2025,
        month = feb,
       volume = {979},
       number = {2},
          eid = {L41},
        pages = {L41},
          doi = {10.3847/2041-8213/adabc2},
archivePrefix = {arXiv},
       eprint = {2411.15546},
 primaryClass = {astro-ph.HE},
       adsurl = {https://ui.adsabs.harvard.edu/abs/2025ApJ...979L..41L}
}

@ARTICLE{Zhang2020,
       author = {{Zhang}, G.~Q. and {Yu}, Hai and {He}, J.~H. and {Wang}, F.~Y.},
        title = "{Dispersion Measures of Fast Radio Burst Host Galaxies Derived from IllustrisTNG Simulation}",
      journal = {\apj},
         year = 2020,
        month = sep,
       volume = {900},
       number = {2},
          eid = {170},
        pages = {170},
          doi = {10.3847/1538-4357/abaa4a},
archivePrefix = {arXiv},
       eprint = {2007.13935},
 primaryClass = {astro-ph.HE},
       adsurl = {https://ui.adsabs.harvard.edu/abs/2020ApJ...900..170Z}
}

@ARTICLE{1,
       author = {{Merritt}, David and {Ferrarese}, Laura},
        title = "{The M$_{{\textbullet}}$-{\ensuremath{\sigma}} Relation for Supermassive Black Holes}",
      journal = {\apj},
         year = 2001,
        month = jan,
       volume = {547},
       number = {1},
        pages = {140-145},
          doi = {10.1086/318372},
archivePrefix = {arXiv},
       eprint = {astro-ph/0008310},
 primaryClass = {astro-ph},
       adsurl = {https://ui.adsabs.harvard.edu/abs/2001ApJ...547..140M}
}

@ARTICLE{2,
       author = {{Aller}, M.~C. and {Richstone}, D.},
        title = "{The Cosmic Density of Massive Black Holes from Galaxy Velocity Dispersions}",
      journal = {\aj},
         year = 2002,
        month = dec,
       volume = {124},
       number = {6},
        pages = {3035-3041},
          doi = {10.1086/344484},
archivePrefix = {arXiv},
       eprint = {astro-ph/0210573},
 primaryClass = {astro-ph},
       adsurl = {https://ui.adsabs.harvard.edu/abs/2002AJ....124.3035A}
}

@ARTICLE{3,
       author = {{Freitag}, Marc},
        title = "{Monte Carlo cluster simulations to determine the rate of compact star inspiralling to a central galactic black hole}",
      journal = {Classical and Quantum Gravity},
         year = 2001,
        month = oct,
       volume = {18},
       number = {19},
        pages = {4033-4038},
          doi = {10.1088/0264-9381/18/19/309},
archivePrefix = {arXiv},
       eprint = {astro-ph/0107193},
 primaryClass = {astro-ph},
       adsurl = {https://ui.adsabs.harvard.edu/abs/2001CQGra..18.4033F}
}

@ARTICLE{Zhang2018b,
       author = {{Zhang}, Bing},
        title = "{A {\textquotedblleft}Cosmic Comb{\textquotedblright} Model of Fast Radio Bursts}",
      journal = {\apjl},
         year = 2017,
        month = feb,
       volume = {836},
       number = {2},
          eid = {L32},
        pages = {L32},
          doi = {10.3847/2041-8213/aa5ded},
archivePrefix = {arXiv},
       eprint = {1701.04094},
 primaryClass = {astro-ph.HE},
       adsurl = {https://ui.adsabs.harvard.edu/abs/2017ApJ...836L..32Z}
}

@string{june = {June}}

@article{chime_cat,
 adsurl = {https://ui.adsabs.harvard.edu/abs/2021arXiv210604352T},
 archiveprefix = {arXiv},
 author = {{The CHIME/FRB Collaboration} and {:} and {Amiri}, Mandana and {Andersen}, Bridget C. and {Bandura}, Kevin and {Berger}, Sabrina and {Bhardwaj}, Mohit and {Boyce}, Michelle M. and {Boyle}, P.~J. and {Brar}, Charanjot and {Breitman}, Daniela and {Cassanelli}, Tomas and {Chawla}, Pragya and {Chen}, Tianyue and {Cliche}, J. -F. and {Cook}, Amanda and {Cubranic}, Davor and {Curtin}, Alice P. and {Deng}, Meiling and {Dobbs}, Matt and {Fengqiu} and {Dong} and {Eadie}, Gwendolyn and {Fandino}, Mateus and {Fonseca}, Emmanuel and {Gaensler}, B.~M. and {Giri}, Utkarsh and {Good}, Deborah C. and {Halpern}, Mark and {Hill}, Alex S. and {Hinshaw}, Gary and {Josephy}, Alexander and {Kaczmarek}, Jane F. and {Kader}, Zarif and {Kania}, Joseph W. and {Kaspi}, Victoria M. and {Landecker}, T.~L. and {Lang}, Dustin and {Leung}, Calvin and {Li}, Dongzi and {Lin}, Hsiu-Hsien and {Masui}, Kiyoshi W. and {Mckinven}, Ryan and {Mena-Parra}, Juan and {Merryfield}, Marcus and {Meyers}, Bradley W. and {Michilli}, Daniele and {Milutinovic}, Nikola and {Mirhosseini}, Arash and {M{\"u}nchmeyer}, Moritz and {Naidu}, Arun and {Newburgh}, Laura and {Ng}, Cherry and {Patel}, Chitrang and {Pen}, Ue-Li and {Petroff}, Emily and {Pinsonneault-Marotte}, Tristan and {Pleunis}, Ziggy and {Rafiei-Ravandi}, Masoud and {Rahman}, Mubdi and {Ransom}, Scott M. and {Renard}, Andre and {Sanghavi}, Pranav and {Scholz}, Paul and {Shaw}, J. Richard and {Shin}, Kaitlyn and {Siegel}, Seth R. and {Sikora}, Andrew E. and {Singh}, Saurabh and {Smith}, Kendrick M. and {Stairs}, Ingrid and {Tan}, Chia Min and {Tendulkar}, S.~P. and {Vanderlinde}, Keith and {Wang}, Haochen and {Wulf}, Dallas and {Zwaniga}, A.~V.},
 eid = {arXiv:2106.04352},
 eprint = {2106.04352},
 journal = {arXiv e-prints},
 month = {June},
 pages = {arXiv:2106.04352},
 primaryclass = {astro-ph.HE},
 title = {{The First CHIME/FRB Fast Radio Burst Catalog}},
 year = {2021}
}

@article{Lorimer2007,
 adsurl = {https://ui.adsabs.harvard.edu/abs/2007Sci...318..777L},
 archiveprefix = {arXiv},
 author = {{Lorimer}, D.~R. and {Bailes}, M. and {McLaughlin}, M.~A. and {Narkevic}, D.~J. and {Crawford}, F.},
 doi = {10.1126/science.1147532},
 eprint = {0709.4301},
 journal = {Science},
 month = {November},
 number = {5851},
 pages = {777},
 primaryclass = {astro-ph},
 title = {{A Bright Millisecond Radio Burst of Extragalactic Origin}},
 volume = {318},
 year = {2007}
}

@ARTICLE{ne2001,
       author = {{Cordes}, J.~M. and {Lazio}, T.~J.~W.},
        title = "{NE2001.I. A New Model for the Galactic Distribution of Free Electrons and its Fluctuations}",
      journal = {arXiv e-prints},
         year = 2002,
        month = jul,
          eid = {astro-ph/0207156},
        pages = {astro-ph/0207156},
          doi = {10.48550/arXiv.astro-ph/0207156},
archivePrefix = {arXiv},
       eprint = {astro-ph/0207156},
 primaryClass = {astro-ph},
       adsurl = {https://ui.adsabs.harvard.edu/abs/2002astro.ph..7156C}
}

@article{Planck2020,
 adsurl = {https://ui.adsabs.harvard.edu/abs/2020A&A...641A...6P},
 archiveprefix = {arXiv},
 author = {{Planck Collaboration} and {Aghanim}, N. and {Akrami}, Y. and {Ashdown}, M. and {Aumont}, J. and {Baccigalupi}, C. and {Ballardini}, M. and {Banday}, A.~J. and {Barreiro}, R.~B. and {Bartolo}, N. and {Basak}, S. and {Battye}, R. and {Benabed}, K. and {Bernard}, J. -P. and {Bersanelli}, M. and {Bielewicz}, P. and {Bock}, J.~J. and {Bond}, J.~R. and {Borrill}, J. and {Bouchet}, F.~R. and {Boulanger}, F. and {Bucher}, M. and {Burigana}, C. and {Butler}, R.~C. and {Calabrese}, E. and {Cardoso}, J. -F. and {Carron}, J. and {Challinor}, A. and {Chiang}, H.~C. and {Chluba}, J. and {Colombo}, L.~P.~L. and {Combet}, C. and {Contreras}, D. and {Crill}, B.~P. and {Cuttaia}, F. and {de Bernardis}, P. and {de Zotti}, G. and {Delabrouille}, J. and {Delouis}, J. -M. and {Di Valentino}, E. and {Diego}, J.~M. and {Dor{\'e}}, O. and {Douspis}, M. and {Ducout}, A. and {Dupac}, X. and {Dusini}, S. and {Efstathiou}, G. and {Elsner}, F. and {En{\ss}lin}, T.~A. and {Eriksen}, H.~K. and {Fantaye}, Y. and {Farhang}, M. and {Fergusson}, J. and {Fernandez-Cobos}, R. and {Finelli}, F. and {Forastieri}, F. and {Frailis}, M. and {Fraisse}, A.~A. and {Franceschi}, E. and {Frolov}, A. and {Galeotta}, S. and {Galli}, S. and {Ganga}, K. and {G{\'e}nova-Santos}, R.~T. and {Gerbino}, M. and {Ghosh}, T. and {Gonz{\'a}lez-Nuevo}, J. and {G{\'o}rski}, K.~M. and {Gratton}, S. and {Gruppuso}, A. and {Gudmundsson}, J.~E. and {Hamann}, J. and {Handley}, W. and {Hansen}, F.~K. and {Herranz}, D. and {Hildebrandt}, S.~R. and {Hivon}, E. and {Huang}, Z. and {Jaffe}, A.~H. and {Jones}, W.~C. and {Karakci}, A. and {Keih{\"a}nen}, E. and {Keskitalo}, R. and {Kiiveri}, K. and {Kim}, J. and {Kisner}, T.~S. and {Knox}, L. and {Krachmalnicoff}, N. and {Kunz}, M. and {Kurki-Suonio}, H. and {Lagache}, G. and {Lamarre}, J. -M. and {Lasenby}, A. and {Lattanzi}, M. and {Lawrence}, C.~R. and {Le Jeune}, M. and {Lemos}, P. and {Lesgourgues}, J. and {Levrier}, F. and {Lewis}, A. and {Liguori}, M. and {Lilje}, P.~B. and {Lilley}, M. and {Lindholm}, V. and {L{\'o}pez-Caniego}, M. and {Lubin}, P.~M. and {Ma}, Y. -Z. and {Mac{\'\i}as-P{\'e}rez}, J.~F. and {Maggio}, G. and {Maino}, D. and {Mandolesi}, N. and {Mangilli}, A. and {Marcos-Caballero}, A. and {Maris}, M. and {Martin}, P.~G. and {Martinelli}, M. and {Mart{\'\i}nez-Gonz{\'a}lez}, E. and {Matarrese}, S. and {Mauri}, N. and {McEwen}, J.~D. and {Meinhold}, P.~R. and {Melchiorri}, A. and {Mennella}, A. and {Migliaccio}, M. and {Millea}, M. and {Mitra}, S. and {Miville-Desch{\^e}nes}, M. -A. and {Molinari}, D. and {Montier}, L. and {Morgante}, G. and {Moss}, A. and {Natoli}, P. and {N{\o}rgaard-Nielsen}, H.~U. and {Pagano}, L. and {Paoletti}, D. and {Partridge}, B. and {Patanchon}, G. and {Peiris}, H.~V. and {Perrotta}, F. and {Pettorino}, V. and {Piacentini}, F. and {Polastri}, L. and {Polenta}, G. and {Puget}, J. -L. and {Rachen}, J.~P. and {Reinecke}, M. and {Remazeilles}, M. and {Renzi}, A. and {Rocha}, G. and {Rosset}, C. and {Roudier}, G. and {Rubi{\~n}o-Mart{\'\i}n}, J.~A. and {Ruiz-Granados}, B. and {Salvati}, L. and {Sandri}, M. and {Savelainen}, M. and {Scott}, D. and {Shellard}, E.~P.~S. and {Sirignano}, C. and {Sirri}, G. and {Spencer}, L.~D. and {Sunyaev}, R. and {Suur-Uski}, A. -S. and {Tauber}, J.~A. and {Tavagnacco}, D. and {Tenti}, M. and {Toffolatti}, L. and {Tomasi}, M. and {Trombetti}, T. and {Valenziano}, L. and {Valiviita}, J. and {Van Tent}, B. and {Vibert}, L. and {Vielva}, P. and {Villa}, F. and {Vittorio}, N. and {Wandelt}, B.~D. and {Wehus}, I.~K. and {White}, M. and {White}, S.~D.~M. and {Zacchei}, A. and {Zonca}, A.},
 doi = {10.1051/0004-6361/201833910},
 eid = {A6},
 eprint = {1807.06209},
 journal = {AAP},
 month = {September},
 pages = {A6},
 primaryclass = {astro-ph.CO},
 title = {{Planck 2018 results. VI. Cosmological parameters}},
 volume = {641},
 year = {2020}
}

@article{Spitler2014,
 author = {L. G. Spitler and J. M. Cordes and J. W. T. Hessels and D. R. Lorimer and M. A. McLaughlin and S. Chatterjee and F. Crawford and J. S. Deneva and V. M. Kaspi and R. S. Wharton and B. Allen and S. Bogdanov and A. Brazier and F. Camilo and P. C. C. Freire and F. A. Jenet and C. Karako-Argaman and B. Knispel and P. Lazarus and K. J. Lee and J. van Leeuwen and R. Lynch and S. M. Ransom and P. Scholz and X. Siemens and I. H. Stairs and K. Stovall and J. K. Swiggum and A. Venkataraman and W. W. Zhu and C. Aulbert and H. Fehrmann},
 doi = {10.1088/0004-637x/790/2/101},
 journal = {The Astrophysical Journal},
 month = {July},
 number = {2},
 pages = {101},
 publisher = {{IOP} Publishing},
 title = {{FAST} {RADIO} {BURST} {DISCOVERED} {IN} {THE} {ARECIBO} {PULSAR} {ALFA} {SURVEY}},
 url = {https://doi.org/10.1088/0004-637x/790/2/101},
 volume = {790},
 year = {2014}
}

@ARTICLE{Xiao2021,
       author = {{Xiao}, Di and {Wang}, FaYin and {Dai}, ZiGao},
        title = "{The physics of fast radio bursts}",
      journal = {Science China Physics, Mechanics, and Astronomy},
         year = 2021,
        month = apr,
       volume = {64},
       number = {4},
          eid = {249501},
        pages = {249501},
          doi = {10.1007/s11433-020-1661-7},
archivePrefix = {arXiv},
       eprint = {2101.04907},
 primaryClass = {astro-ph.HE},
       adsurl = {https://ui.adsabs.harvard.edu/abs/2021SCPMA..6449501X}
}

@ARTICLE{Wang2022,
       author = {{Wang}, F.~Y. and {Zhang}, G.~Q. and {Dai}, Z.~G. and {Cheng}, K.~S.},
        title = "{Repeating fast radio burst 20201124A originates from a magnetar/Be star binary}",
      journal = {Nature Communications},
         year = 2022,
        month = sep,
       volume = {13},
          eid = {4382},
        pages = {4382},
          doi = {10.1038/s41467-022-31923-y},
archivePrefix = {arXiv},
       eprint = {2204.08124},
 primaryClass = {astro-ph.HE},
       adsurl = {https://ui.adsabs.harvard.edu/abs/2022NatCo..13.4382W}
}

@ARTICLE{Prochaska2019,
       author = {{Prochaska}, J. Xavier and {Macquart}, Jean-Pierre and
         {McQuinn}, Matthew and {Simha}, Sunil and {Shannon}, Ryan M. and
         {Day}, Cherie K. and {Marnoch}, Lachlan and {Ryder}, Stuart and
         {Deller}, Adam and {Bannister}, Keith W. and {Bhandari}, Shivani and
         {Bordoloi}, Rongmon and {Bunton}, John and {Cho}, Hyerin and
         {Flynn}, Chris and {Mahony}, Elizabeth K. and {Phillips}, Chris and
         {Qiu}, Hao and {Tejos}, Nicolas},
        title = "{The low density and magnetization of a massive galaxy halo exposed by a fast radio burst}",
      journal = {Science},
         year = 2019,
        month = oct,
       volume = {366},
       number = {6462},
        pages = {231-234},
          doi = {10.1126/science.aay0073},
archivePrefix = {arXiv},
       eprint = {1909.11681},
 primaryClass = {astro-ph.GA},
       adsurl = {https://ui.adsabs.harvard.edu/abs/2019Sci...366..231P}
}

@ARTICLE{Macquart2020,
       author = {{Macquart}, J. -P. and {Prochaska}, J.~X. and {McQuinn}, M. and
         {Bannister}, K.~W. and {Bhandari}, S. and {Day}, C.~K. and
         {Deller}, A.~T. and {Ekers}, R.~D. and {James}, C.~W. and
         {Marnoch}, L. and {Os{\l}owski}, S. and {Phillips}, C. and
         {Ryder}, S.~D. and {Scott}, D.~R. and {Shannon}, R.~M. and {Tejos}, N.},
        title = "{A census of baryons in the Universe from localized fast radio bursts}",
      journal = {\nat},
         year = 2020,
        month = may,
       volume = {581},
       number = {7809},
        pages = {391-395},
          doi = {10.1038/s41586-020-2300-2},
archivePrefix = {arXiv},
       eprint = {2005.13161},
 primaryClass = {astro-ph.CO},
       adsurl = {https://ui.adsabs.harvard.edu/abs/2020Natur.581..391M}
}

@ARTICLE{Bochenek2020,
       author = {{Bochenek}, Christopher D. and {Ravi}, Vikram and
         {Belov}, Konstantin V. and {Hallinan}, Gregg and {Kocz}, Jonathon and
         {Kulkarni}, Shri R. and {McKenna}, Dan L.},
        title = "{A fast radio burst associated with a Galactic magnetar}",
      journal = {arXiv e-prints},
         year = 2020,
        month = may,
          eid = {arXiv:2005.10828},
        pages = {arXiv:2005.10828},
archivePrefix = {arXiv},
       eprint = {2005.10828},
 primaryClass = {astro-ph.HE},
       adsurl = {https://ui.adsabs.harvard.edu/abs/2020arXiv200510828B}
}

@ARTICLE{Michilli2018,
       author = {{Michilli}, D. and {Seymour}, A. and {Hessels}, J.~W.~T. and
         {Spitler}, L.~G. and {Gajjar}, V. and {Archibald}, A.~M. and
         {Bower}, G.~C. and {Chatterjee}, S. and {Cordes}, J.~M. and
         {Gourdji}, K. and {Heald}, G.~H. and {Kaspi}, V.~M. and {Law}, C.~J. and
         {Sobey}, C. and {Adams}, E.~A.~K. and {Bassa}, C.~G. and
         {Bogdanov}, S. and {Brinkman}, C. and {Demorest}, P. and {Fernand
        ez}, F. and {Hellbourg}, G. and {Lazio}, T.~J.~W. and {Lynch}, R.~S. and
         {Maddox}, N. and {Marcote}, B. and {McLaughlin}, M.~A. and
         {Paragi}, Z. and {Ransom}, S.~M. and {Scholz}, P. and
         {Siemion}, A.~P.~V. and {Tendulkar}, S.~P. and {van Rooy}, P. and
         {Wharton}, R.~S. and {Whitlow}, D.},
        title = "{An extreme magneto-ionic environment associated with the fast radio burst source FRB 121102}",
      journal = {\nat},
         year = 2018,
        month = jan,
       volume = {553},
       number = {7687},
        pages = {182-185},
          doi = {10.1038/nature25149},
archivePrefix = {arXiv},
       eprint = {1801.03965},
 primaryClass = {astro-ph.HE},
       adsurl = {https://ui.adsabs.harvard.edu/abs/2018Natur.553..182M}
}

@ARTICLE{CHIME/FRB2020,
       author = {{The CHIME/FRB Collaboration} and {:} and {Andersen}, B.~C. and {Band
        ura}, K.~M. and {Bhardwaj}, M. and {Bij}, A. and {Boyce}, M.~M. and
         {Boyle}, P.~J. and {Brar}, C. and {Cassanelli}, T. and {Chawla}, P. and
         {Chen}, T. and {Cliche}, J. -F. and {Cook}, A. and {Cubranic}, D. and
         {Curtin}, A.~P. and {Denman}, N.~T. and {Dobbs}, M. and {Dong}, F.~Q. and
         {Fandino}, M. and {Fonseca}, E. and {Gaensler}, B.~M. and {Giri}, U. and
         {Good}, D.~C. and {Halpern}, M. and {Hill}, A.~S. and {Hinshaw}, G.~F. and
         {H{\"o}fer}, C. and {Josephy}, A. and {Kania}, J.~W. and
         {Kaspi}, V.~M. and {Landecker}, T.~L. and {Leung}, C. and {Li}, D.~Z. and
         {Lin}, H. -H. and {Masui}, K.~W. and {Mckinven}, R. and
         {Mena-Parra}, J. and {Merryfield}, M. and {Meyers}, B.~W. and
         {Michilli}, D. and {Milutinovic}, N. and {Mirhosseini}, A. and
         {M{\"u}nchmeyer}, M. and {Naidu}, A. and {Newburgh}, L.~B. and
         {Ng}, C. and {Patel}, C. and {Pen}, U. -L. and
         {Pinsonneault-Marotte}, T. and {Pleunis}, Z. and {Quine}, B.~M. and
         {Rafiei-Ravandi}, M. and {Rahman}, M. and {Ransom}, S.~M. and
         {Renard}, A. and {Sanghavi}, P. and {Scholz}, P. and {Shaw}, J.~R. and
         {Shin}, K. and {Siegel}, S.~R. and {Singh}, S. and {Smegal}, R.~J. and
         {Smith}, K.~M. and {Stairs}, I.~H. and {Tan}, C.~M. and
         {Tendulkar}, S.~P. and {Tretyakov}, I. and {Vanderlinde}, K. and
         {Wang}, H. and {Wulf}, D. and {Zwaniga}, A.~V.},
        title = "{A bright millisecond-duration radio burst from a Galactic magnetar}",
      journal = {arXiv e-prints},
         year = 2020,
        month = may,
          eid = {arXiv:2005.10324},
        pages = {arXiv:2005.10324},
archivePrefix = {arXiv},
       eprint = {2005.10324},
 primaryClass = {astro-ph.HE},
       adsurl = {https://ui.adsabs.harvard.edu/abs/2020arXiv200510324T}
}

@ARTICLE{Wu2022,
       author = {{Wu}, Qin and {Zhang}, Guo-Qiang and {Wang}, Fa-Yin},
        title = "{An 8 per cent determination of the Hubble constant from localized fast radio bursts}",
      journal = {\mnras},
         year = 2022,
        month = sep,
       volume = {515},
       number = {1},
        pages = {L1-L5},
          doi = {10.1093/mnrasl/slac022},
archivePrefix = {arXiv},
       eprint = {2108.00581},
 primaryClass = {astro-ph.CO},
       adsurl = {https://ui.adsabs.harvard.edu/abs/2022MNRAS.515L...1W}
}

@ARTICLE{Xu2020,
       author = {{Xu}, Siyao and {Zhang}, Bing},
        title = "{Probing the Intergalactic Turbulence with Fast Radio Bursts}",
      journal = {\apjl},
         year = 2020,
        month = aug,
       volume = {898},
       number = {2},
          eid = {L48},
        pages = {L48},
          doi = {10.3847/2041-8213/aba760},
archivePrefix = {arXiv},
       eprint = {2007.04089},
 primaryClass = {astro-ph.HE},
       adsurl = {https://ui.adsabs.harvard.edu/abs/2020ApJ...898L..48X}
}

@ARTICLE{Xu2016,
       author = {{Xu}, Siyao and {Zhang}, Bing},
        title = "{Interpretation of the Structure Function of Rotation Measure in the Interstellar Medium}",
      journal = {\apj},
         year = 2016,
        month = jun,
       volume = {824},
       number = {2},
          eid = {113},
        pages = {113},
          doi = {10.3847/0004-637X/824/2/113},
archivePrefix = {arXiv},
       eprint = {1604.05445},
 primaryClass = {astro-ph.GA},
       adsurl = {https://ui.adsabs.harvard.edu/abs/2016ApJ...824..113X}
}

@ARTICLE{Elmegreen2004,
       author = {{Elmegreen}, Bruce G. and {Scalo}, John},
        title = "{Interstellar Turbulence I: Observations and Processes}",
      journal = {\araa},
         year = 2004,
        month = sep,
       volume = {42},
       number = {1},
        pages = {211-273},
          doi = {10.1146/annurev.astro.41.011802.094859},
archivePrefix = {arXiv},
       eprint = {astro-ph/0404451},
 primaryClass = {astro-ph},
       adsurl = {https://ui.adsabs.harvard.edu/abs/2004ARA&A..42..211E}
}

@ARTICLE{Lazarian2016,
       author = {{Lazarian}, A. and {Pogosyan}, D.},
        title = "{Spectrum and Anisotropy of Turbulence from Multi-frequency Measurement of Synchrotron Polarization}",
      journal = {\apj},
         year = 2016,
        month = feb,
       volume = {818},
       number = {2},
          eid = {178},
        pages = {178},
          doi = {10.3847/0004-637X/818/2/178},
archivePrefix = {arXiv},
       eprint = {1511.01537},
 primaryClass = {astro-ph.GA},
       adsurl = {https://ui.adsabs.harvard.edu/abs/2016ApJ...818..178L}
}

@ARTICLE{Xu2021,
       author = {{Xu}, Siyao and {Weinberg}, David H. and {Zhang}, Bing},
        title = "{Statistical Measurements of Dispersion Measure Fluctuations in Fast Radio Bursts}",
      journal = {\apjl},
         year = 2021,
        month = dec,
       volume = {922},
       number = {2},
          eid = {L31},
        pages = {L31},
          doi = {10.3847/2041-8213/ac399c},
archivePrefix = {arXiv},
       eprint = {2111.07417},
 primaryClass = {astro-ph.CO},
       adsurl = {https://ui.adsabs.harvard.edu/abs/2021ApJ...922L..31X}
}

@BOOK{Biskamp2003,
       author = {{Biskamp}, Dieter},
        title = "{Magnetohydrodynamic Turbulence}",
         year = 2003,
       adsurl = {https://ui.adsabs.harvard.edu/abs/2003matu.book.....B}
}

@ARTICLE{Ryu2008,
       author = {{Ryu}, Dongsu and {Kang}, Hyesung and {Cho}, Jungyeon and {Das}, Santabrata},
        title = "{Turbulence and Magnetic Fields in the Large-Scale Structure of the Universe}",
      journal = {Science},
         year = 2008,
        month = may,
       volume = {320},
       number = {5878},
        pages = {909},
          doi = {10.1126/science.1154923},
archivePrefix = {arXiv},
       eprint = {0805.2466},
 primaryClass = {astro-ph},
       adsurl = {https://ui.adsabs.harvard.edu/abs/2008Sci...320..909R}
}

@ARTICLE{Connor2024,
       author = {{Connor}, Liam and {Ravi}, Vikram and {Sharma}, Kritti and {Ocker}, Stella Koch and {Faber}, Jakob and {Hallinan}, Gregg and {Harnach}, Charlie and {Hellbourg}, Greg and {Hobbs}, Rick and {Hodge}, David and {Hodges}, Mark and {Kosogorov}, Nikita and {Lamb}, James and {Law}, Casey and {Rasmussen}, Paul and {Sherman}, Myles and {Somalwar}, Jean and {Weinreb}, Sander and {Woody}, David},
        title = "{A gas rich cosmic web revealed by partitioning the missing baryons}",
      journal = {arXiv e-prints},
         year = 2024,
        month = sep,
          eid = {arXiv:2409.16952},
        pages = {arXiv:2409.16952},
          doi = {10.48550/arXiv.2409.16952},
archivePrefix = {arXiv},
       eprint = {2409.16952},
 primaryClass = {astro-ph.CO},
       adsurl = {https://ui.adsabs.harvard.edu/abs/2024arXiv240916952C}
}

@ARTICLE{McQuinn2014,
       author = {{McQuinn}, Matthew},
        title = "{Locating the ``Missing'' Baryons with Extragalactic Dispersion Measure Estimates}",
      journal = {\apjl},
         year = 2014,
        month = jan,
       volume = {780},
       number = {2},
          eid = {L33},
        pages = {L33},
          doi = {10.1088/2041-8205/780/2/L33},
archivePrefix = {arXiv},
       eprint = {1309.4451},
 primaryClass = {astro-ph.CO},
       adsurl = {https://ui.adsabs.harvard.edu/abs/2014ApJ...780L..33M}
}

@ARTICLE{Zhang2023Review,
       author = {{Zhang}, Bing},
        title = "{The physics of fast radio bursts}",
      journal = {Reviews of Modern Physics},
         year = 2023,
        month = jul,
       volume = {95},
       number = {3},
          eid = {035005},
        pages = {035005},
          doi = {10.1103/RevModPhys.95.035005},
archivePrefix = {arXiv},
       eprint = {2212.03972},
 primaryClass = {astro-ph.HE},
       adsurl = {https://ui.adsabs.harvard.edu/abs/2023RvMP...95c5005Z}
}

@ARTICLE{ZhangZJ2021,
       author = {{Zhang}, Z.~J. and {Yan}, K. and {Li}, C.~M. and {Zhang}, G.~Q. and {Wang}, F.~Y.},
        title = "{Intergalactic Medium Dispersion Measures of Fast Radio Bursts Estimated from IllustrisTNG Simulation and Their Cosmological Applications}",
      journal = {\apj},
         year = 2021,
        month = jan,
       volume = {906},
       number = {1},
          eid = {49},
        pages = {49},
          doi = {10.3847/1538-4357/abceb9},
archivePrefix = {arXiv},
       eprint = {2011.14494},
 primaryClass = {astro-ph.CO},
       adsurl = {https://ui.adsabs.harvard.edu/abs/2021ApJ...906...49Z}
}

@ARTICLE{YangKB2022,
       author = {{Yang}, K.~B. and {Wu}, Q. and {Wang}, F.~Y.},
        title = "{Finding the Missing Baryons in the Intergalactic Medium with Localized Fast Radio Bursts}",
      journal = {\apjl},
         year = 2022,
        month = dec,
       volume = {940},
       number = {2},
          eid = {L29},
        pages = {L29},
          doi = {10.3847/2041-8213/aca145},
archivePrefix = {arXiv},
       eprint = {2211.04058},
 primaryClass = {astro-ph.HE},
       adsurl = {https://ui.adsabs.harvard.edu/abs/2022ApJ...940L..29Y}
}

@ARTICLE{Deng2014,
       author = {{Deng}, Wei and {Zhang}, Bing},
        title = "{Cosmological Implications of Fast Radio Burst/Gamma-Ray Burst Associations}",
      journal = {\apjl},
         year = 2014,
        month = mar,
       volume = {783},
       number = {2},
          eid = {L35},
        pages = {L35},
          doi = {10.1088/2041-8205/783/2/L35},
archivePrefix = {arXiv},
       eprint = {1401.0059},
 primaryClass = {astro-ph.HE},
       adsurl = {https://ui.adsabs.harvard.edu/abs/2014ApJ...783L..35D}
}

@ARTICLE{ZhangZ2025,
       author = {{Zhang}, Zhao and {Nagamine}, Kentaro and {Oku}, Yuri and {Lee}, Khee-Gan and {Fukushima}, Keita and {Tomaru}, Kazuki and {Zhang}, Bing and {Medlock}, Isabel and {Nagai}, Daisuke},
        title = "{Probing the cosmic baryon distribution and the impact of AGN feedback with FRBs in CROCODILE simulation}",
      journal = {arXiv e-prints},
         year = 2025,
        month = mar,
          eid = {arXiv:2503.12741},
        pages = {arXiv:2503.12741},
          doi = {10.48550/arXiv.2503.12741},
archivePrefix = {arXiv},
       eprint = {2503.12741},
 primaryClass = {astro-ph.CO},
       adsurl = {https://ui.adsabs.harvard.edu/abs/2025arXiv250312741Z}
}

@ARTICLE{Walker2024,
       author = {{Walker}, Charles R.~H. and {Spitler}, Laura G. and {Ma}, Yin-Zhe and {Cheng}, Cheng and {Artale}, Maria Celeste and {Hummels}, Cameron B.},
        title = "{The dispersion measure contributions of the cosmic web}",
      journal = {\aap},
         year = 2024,
        month = mar,
       volume = {683},
          eid = {A71},
        pages = {A71},
          doi = {10.1051/0004-6361/202347139},
archivePrefix = {arXiv},
       eprint = {2309.08268},
 primaryClass = {astro-ph.CO},
       adsurl = {https://ui.adsabs.harvard.edu/abs/2024A&A...683A..71W}
}

@ARTICLE{Batten2021,
       author = {{Batten}, Adam J. and {Duffy}, Alan R. and {Wijers}, Nastasha A. and {Gupta}, Vivek and {Flynn}, Chris and {Schaye}, Joop and {Ryan-Weber}, Emma},
        title = "{The cosmic dispersion measure in the EAGLE simulations}",
      journal = {\mnras},
         year = 2021,
        month = aug,
       volume = {505},
       number = {4},
        pages = {5356-5369},
          doi = {10.1093/mnras/stab1528},
archivePrefix = {arXiv},
       eprint = {2011.14547},
 primaryClass = {astro-ph.CO},
       adsurl = {https://ui.adsabs.harvard.edu/abs/2021MNRAS.505.5356B}
}

@ARTICLE{Xu2016_scattering,
       author = {{Xu}, Siyao and {Zhang}, Bing},
        title = "{On the Origin of the Scatter Broadening of Fast Radio Burst Pulses and Astrophysical Implications}",
      journal = {\apj},
         year = 2016,
        month = dec,
       volume = {832},
       number = {2},
          eid = {199},
        pages = {199},
          doi = {10.3847/0004-637X/832/2/199},
archivePrefix = {arXiv},
       eprint = {1608.03930},
 primaryClass = {astro-ph.HE},
       adsurl = {https://ui.adsabs.harvard.edu/abs/2016ApJ...832..199X}
}

@ARTICLE{Zhu2010,
       author = {{Zhu}, Weishan and {Feng}, Long-long and {Fang}, Li-Zhi},
        title = "{Vorticity of Intergalactic Medium Velocity Field on Large Scales}",
      journal = {\apj},
         year = 2010,
        month = mar,
       volume = {712},
       number = {1},
        pages = {1-13},
          doi = {10.1088/0004-637X/712/1/1},
archivePrefix = {arXiv},
       eprint = {1001.4127},
 primaryClass = {astro-ph.CO},
       adsurl = {https://ui.adsabs.harvard.edu/abs/2010ApJ...712....1Z}
}

@ARTICLE{Ioka2003,
       author = {{Ioka}, Kunihito},
        title = "{The Cosmic Dispersion Measure from Gamma-Ray Burst Afterglows: Probing the Reionization History and the Burst Environment}",
      journal = {\apjl},
         year = 2003,
        month = dec,
       volume = {598},
       number = {2},
        pages = {L79-L82},
          doi = {10.1086/380598},
archivePrefix = {arXiv},
       eprint = {astro-ph/0309200},
 primaryClass = {astro-ph},
       adsurl = {https://ui.adsabs.harvard.edu/abs/2003ApJ...598L..79I}
}

@ARTICLE{chimecatalog2,
       author = {{FRB Collaboration} and {Abbott}, Thomas and {Andersen}, Bridget C. and {Andrew}, Shion and {Bandura}, Kevin and {Bhardwaj}, Mohit and {Bhusare}, Yash and {Brar}, Charanjot and {Cassanelli}, Tomas and {Chatterjee}, Shami and {Cliche}, Jean-Francois and {Cook}, Amanda M. and {Curtin}, Alice and {Dobbs}, Matt and {Dong}, Fengqiu Adam and {Eadie}, Gwendolyn and {Eftekhari}, Tarraneh and {Fonseca}, Emmanuel and {Gaensler}, B.~M. and {Good}, Deborah and {Halpern}, Mark and {Hessels}, Jason W.~T. and {Ibik}, Adaeze and {Jain}, Naman and {Joseph}, Ronniy C. and {Kader}, Zarif and {Kaspi}, Victoria M. and {Khan}, Afrokk and {Kharel}, Bikash and {Kumar}, Ajay and {Landecker}, T.~L. and {Lang}, Dustin and {Lanman}, Adam E. and {L'Argent}, Magnus and {Lazda}, Mattias and {Leung}, Calvin and {Li}, Dong Zi and {Lintott}, Chris J. and {Main}, Robert and {Masui}, Kiyoshi W. and {Mate}, Sujay and {McGregor}, Kyle and {Mckinven}, Ryan and {Mena-Parra}, Juan and {Meyers}, Bradley W. and {Michilli}, Daniele and {Ng}, Cherry and {Ng}, Mason and {Nimmo}, Kenzie and {Noble}, Gavin and {Pandhi}, Ayush and {Patil}, Swarali S. and {Pearlman}, Aaron B. and {Pen}, Ue-Li and {Pleunis}, Ziggy and {Prochaska}, J. Xavier and {Rafiei-Ravandi}, Masoud and {Ransom}, Scott and {Renard}, Andre and {Sammons}, Mawson W. and {Sand}, Ketan R. and {Scholz}, Paul and {Shah}, Vishwangi and {Shin}, Kaitlyn and {Siegel}, Seth R. and {Sirota}, Sloane and {Smith}, Kendrick and {Stairs}, Ingrid and {Stenning}, David C. and {Tendulkar}, Shriharsh P. and {Vanderlinde}, Keith and {Walmsley}, Mike and {Wang}, Haochen and {Wulf}, Dallas},
        title = "{The Second CHIME/FRB Catalog of Fast Radio Bursts}",
      journal = {arXiv e-prints},
         year = 2026,
        month = jan,
          eid = {arXiv:2601.09399},
        pages = {arXiv:2601.09399},
          doi = {10.48550/arXiv.2601.09399},
archivePrefix = {arXiv},
       eprint = {2601.09399},
 primaryClass = {astro-ph.HE},
       adsurl = {https://ui.adsabs.harvard.edu/abs/2026arXiv260109399F}
}

@ARTICLE{Armstrong1995,
       author = {{Armstrong}, J.~W. and {Rickett}, B.~J. and {Spangler}, S.~R.},
        title = "{Electron Density Power Spectrum in the Local Interstellar Medium}",
      journal = {\apj},
         year = 1995,
        month = apr,
       volume = {443},
        pages = {209},
          doi = {10.1086/175515},
       adsurl = {https://ui.adsabs.harvard.edu/abs/1995ApJ...443..209A}
}

@ARTICLE{Chepurnov2010,
       author = {{Chepurnov}, A. and {Lazarian}, A. and {Stanimirovi{\'c}}, S. and {Heiles}, Carl and {Peek}, J.~E.~G.},
        title = "{Velocity Spectrum for H I at High Latitudes}",
      journal = {\apj},
         year = 2010,
        month = may,
       volume = {714},
       number = {2},
        pages = {1398-1406},
          doi = {10.1088/0004-637X/714/2/1398},
archivePrefix = {arXiv},
       eprint = {astro-ph/0611462},
 primaryClass = {astro-ph},
       adsurl = {https://ui.adsabs.harvard.edu/abs/2010ApJ...714.1398C}
}

@ARTICLE{Schuecker2004,
       author = {{Schuecker}, P. and {Finoguenov}, A. and {Miniati}, F. and {B{\"o}hringer}, H. and {Briel}, U.~G.},
        title = "{Probing turbulence in the Coma galaxy cluster}",
      journal = {\aap},
         year = 2004,
        month = nov,
       volume = {426},
        pages = {387-397},
          doi = {10.1051/0004-6361:20041039},
archivePrefix = {arXiv},
       eprint = {astro-ph/0404132},
 primaryClass = {astro-ph},
       adsurl = {https://ui.adsabs.harvard.edu/abs/2004A&A...426..387S}
}

@ARTICLE{Vogt2005,
       author = {{Vogt}, C. and {En{\ss}lin}, T.~A.},
        title = "{A Bayesian view on Faraday rotation maps   Seeing the magnetic power spectra in galaxy clusters}",
      journal = {\aap},
         year = 2005,
        month = apr,
       volume = {434},
       number = {1},
        pages = {67-76},
          doi = {10.1051/0004-6361:20041839},
archivePrefix = {arXiv},
       eprint = {astro-ph/0501211},
 primaryClass = {astro-ph},
       adsurl = {https://ui.adsabs.harvard.edu/abs/2005A&A...434...67V}
}

@ARTICLE{Xu2020b,
       author = {{Xu}, Xiaoju and {Zheng}, Zheng},
        title = "{Galaxy assembly bias of central galaxies in the Illustris simulation}",
      journal = {\mnras},
         year = 2020,
        month = feb,
       volume = {492},
       number = {2},
        pages = {2739-2754},
          doi = {10.1093/mnras/staa009},
archivePrefix = {arXiv},
       eprint = {1812.11210},
 primaryClass = {astro-ph.GA},
       adsurl = {https://ui.adsabs.harvard.edu/abs/2020MNRAS.492.2739X}
}

@ARTICLE{Evoli2011,
       author = {{Evoli}, Carmelo and {Ferrara}, Andrea},
        title = "{Turbulence in the intergalactic medium}",
      journal = {\mnras},
         year = 2011,
        month = jun,
       volume = {413},
       number = {4},
        pages = {2721-2734},
          doi = {10.1111/j.1365-2966.2011.18343.x},
archivePrefix = {arXiv},
       eprint = {1101.2449},
 primaryClass = {astro-ph.CO},
       adsurl = {https://ui.adsabs.harvard.edu/abs/2011MNRAS.413.2721E}
}

@ARTICLE{Rorai2017,
       author = {{Rorai}, Alberto and {Hennawi}, Joseph F. and {O{\~n}orbe}, Jose and {White}, Martin and {Prochaska}, J. Xavier and {Kulkarni}, Girish and {Walther}, Michael and {Luki{\'c}}, Zarija and {Lee}, Khee-Gan},
        title = "{Measurement of the small-scale structure of the intergalactic medium using close quasar pairs}",
      journal = {Science},
         year = 2017,
        month = apr,
       volume = {356},
       number = {6336},
        pages = {418-422},
          doi = {10.1126/science.aaf9346},
archivePrefix = {arXiv},
       eprint = {1704.08366},
 primaryClass = {astro-ph.CO},
       adsurl = {https://ui.adsabs.harvard.edu/abs/2017Sci...356..418R}
}

@ARTICLE{Zhu2011,
       author = {{Zhu}, Weishan and {Feng}, Long-Long and {Fang}, Li-Zhi},
        title = "{Dynamical effect of the turbulence of the intergalactic medium on the baryon fraction distribution}",
      journal = {\mnras},
         year = 2011,
        month = aug,
       volume = {415},
       number = {2},
        pages = {1093-1104},
          doi = {10.1111/j.1365-2966.2011.18640.x},
archivePrefix = {arXiv},
       eprint = {1103.1058},
 primaryClass = {astro-ph.CO},
       adsurl = {https://ui.adsabs.harvard.edu/abs/2011MNRAS.415.1093Z}
}

@ARTICLE{Iapichino2011,
       author = {{Iapichino}, L. and {Schmidt}, W. and {Niemeyer}, J.~C. and {Merklein}, J.},
        title = "{Turbulence production and turbulent pressure support in the intergalactic medium}",
      journal = {\mnras},
         year = 2011,
        month = jul,
       volume = {414},
       number = {3},
        pages = {2297-2308},
          doi = {10.1111/j.1365-2966.2011.18550.x},
archivePrefix = {arXiv},
       eprint = {1102.3352},
 primaryClass = {astro-ph.CO},
       adsurl = {https://ui.adsabs.harvard.edu/abs/2011MNRAS.414.2297I}
}

@ARTICLE{Oku2024,
       author = {{Oku}, Yuri and {Nagamine}, Kentaro},
        title = "{Osaka Feedback Model. III. Cosmological Simulation CROCODILE}",
      journal = {\apj},
         year = 2024,
        month = nov,
       volume = {975},
       number = {2},
          eid = {183},
        pages = {183},
          doi = {10.3847/1538-4357/ad77d3},
archivePrefix = {arXiv},
       eprint = {2401.06324},
 primaryClass = {astro-ph.GA},
       adsurl = {https://ui.adsabs.harvard.edu/abs/2024ApJ...975..183O}
}

@ARTICLE{Schulz1981,
       author = {{Schulz-Dubois}, E.~O. and {Rehberg}, I.},
        title = "{Structure function in lieu of correlation function}",
      journal = {Applied Physics},
         year = 1981,
        month = apr,
       volume = {24},
       number = {4},
        pages = {323-329},
          doi = {10.1007/BF00899730},
       adsurl = {https://ui.adsabs.harvard.edu/abs/1981ApPhy..24..323S}
}

@ARTICLE{blinkverse,
       author = {{Xu}, Jiaying and {Feng}, Yi and {Li}, Di and {Wang}, Pei and {Zhang}, Yongkun and {Xie}, Jintao and {Chen}, Huaxi and {Wang}, Han and {Kang}, Zhixuan and {Hu}, Jingjing and {Zheng}, Yun and {Tsai}, Chao-Wei and {Chen}, Xianglei and {Zhou}, Dengke},
        title = "{Blinkverse: A Database of Fast Radio Bursts}",
      journal = {Universe},
         year = 2023,
        month = jul,
       volume = {9},
       number = {7},
          eid = {330},
        pages = {330},
          doi = {10.3390/universe9070330},
archivePrefix = {arXiv},
       eprint = {2308.00336},
 primaryClass = {astro-ph.HE},
       adsurl = {https://ui.adsabs.harvard.edu/abs/2023Univ....9..330X}
}

@ARTICLE{Li2025,
       author = {{Li}, Rui-Nan and {Xu}, Ke and {Gao}, Dao-Hong and {Wu}, Qin and {Yi}, Shuang-Xi and {Wang}, Fa-Yin},
        title = "{Calibrating the DM$_{IGM}${\textendash}z Relation Using Host Galaxy Properties of Fast Radio Bursts}",
      journal = {\apj},
         year = 2025,
        month = aug,
       volume = {989},
       number = {1},
          eid = {77},
        pages = {77},
          doi = {10.3847/1538-4357/adeb72},
       adsurl = {https://ui.adsabs.harvard.edu/abs/2025ApJ...989...77L}
}

@ARTICLE{ymw16,
       author = {{Yao}, J.~M. and {Manchester}, R.~N. and {Wang}, N.},
        title = "{A New Electron-density Model for Estimation of Pulsar and FRB Distances}",
      journal = {\apj},
         year = 2017,
        month = jan,
       volume = {835},
       number = {1},
          eid = {29},
        pages = {29},
          doi = {10.3847/1538-4357/835/1/29},
archivePrefix = {arXiv},
       eprint = {1610.09448},
 primaryClass = {astro-ph.GA},
       adsurl = {https://ui.adsabs.harvard.edu/abs/2017ApJ...835...29Y}
}

@ARTICLE{Marazuela2026,
       author = {{Pastor-Marazuela}, In{\'e}s and {Gordon}, Alexa C. and {Stappers}, Ben and {Khrykin}, Ilya S. and {Tejos}, Nicolas and {Rajwade}, Kaustubh and {Caleb}, Manisha and {Surnis}, Mayuresh P. and {Driessen}, Laura N. and {Simha}, Sunil and {Tian}, Jun and {Prochaska}, J. Xavier and {Barr}, Ewan and {Buchner}, Sarah and {Fong}, Wen-Fai and {Jankowski}, Fabian and {Kahinga}, Lordrick and {Kilpatrick}, Charles D. and {Kramer}, Michael and {Mas-Ribas}, Lluis and {Hennawi}, Joseph},
        title = "{Localization and host galaxy identification of new fast radio bursts with MeerKAT}",
      journal = {\mnras},
         year = 2026,
        month = feb,
       volume = {545},
       number = {4},
          eid = {staf2144},
        pages = {staf2144},
          doi = {10.1093/mnras/staf2144},
archivePrefix = {arXiv},
       eprint = {2507.05982},
 primaryClass = {astro-ph.HE},
       adsurl = {https://ui.adsabs.harvard.edu/abs/2026MNRAS.545f2144P}
}

@ARTICLE{ASKAP,
       author = {{Johnston}, S. and {Taylor}, R. and {Bailes}, M. and {Bartel}, N. and {Baugh}, C. and {Bietenholz}, M. and {Blake}, C. and {Braun}, R. and {Brown}, J. and {Chatterjee}, S. and {Darling}, J. and {Deller}, A. and {Dodson}, R. and {Edwards}, P. and {Ekers}, R. and {Ellingsen}, S. and {Feain}, I. and {Gaensler}, B. and {Haverkorn}, M. and {Hobbs}, G. and {Hopkins}, A. and {Jackson}, C. and {James}, C. and {Joncas}, G. and {Kaspi}, V. and {Kilborn}, V. and {Koribalski}, B. and {Kothes}, R. and {Landecker}, T. and {Lenc}, E. and {Lovell}, J. and {Macquart}, J.-P. and {Manchester}, R. and {Matthews}, D. and {McClure-Griffiths}, N. and {Norris}, R. and {Pen}, U.-L. and {Phillips}, C. and {Power}, C. and {Protheroe}, R. and {Sadler}, E. and {Schmidt}, B. and {Stairs}, I. and {Staveley-Smith}, L. and {Stil}, J. and {Tingay}, S. and {Tzioumis}, A. and {Walker}, M. and {Wall}, J. and {Wolleben}, M.},
        title = "{Science with ASKAP. The Australian square-kilometre-array pathfinder}",
      journal = {Experimental Astronomy},
         year = 2008,
        month = dec,
       volume = {22},
       number = {3},
        pages = {151-273},
          doi = {10.1007/s10686-008-9124-7},
archivePrefix = {arXiv},
       eprint = {0810.5187},
 primaryClass = {astro-ph},
       adsurl = {https://ui.adsabs.harvard.edu/abs/2008ExA....22..151J}
}

@INPROCEEDINGS{MeerKAT,
       author = {{Jonas}, J. and {MeerKAT Team}},
        title = "{The MeerKAT Radio Telescope}",
    booktitle = {MeerKAT Science: On the Pathway to the SKA},
         year = 2016,
        month = jan,
          eid = {1},
        pages = {1},
          doi = {10.22323/1.277.0001},
       adsurl = {https://ui.adsabs.harvard.edu/abs/2016mks..confE...1J}
}

@ARTICLE{SKA,
       author = {{Dewdney}, P.~E. and {Hall}, P.~J. and {Schilizzi}, R.~T. and {Lazio}, T.~J.~L.~W.},
        title = "{The Square Kilometre Array}",
      journal = {IEEE Proceedings},
         year = 2009,
        month = aug,
       volume = {97},
       number = {8},
        pages = {1482-1496},
          doi = {10.1109/JPROC.2009.2021005},
       adsurl = {https://ui.adsabs.harvard.edu/abs/2009IEEEP..97.1482D}
}

@ARTICLE{Niu2026,
       author = {{Niu}, Chen-Hui and {Li}, Di and {Yang}, Yuan-Pei and {Zhu}, Yuhao and {Zhang}, Yongkun and {Zhang}, Jia-Heng and {Du}, Zexin and {Yao}, Jumei and {Zheng}, Xiaoping and {Wang}, Pei and {Feng}, Yi and {Zhang}, Bing and {Zhu}, Weiwei and {Yu}, Wenfei and {Jiang}, Ji-An and {Dai}, Shi and {Tsai}, Chao-Wei and {Chen}, A. Ming and {Hou}, Yijun and {Niu}, Jiarui and {Wang}, Weiyang and {Miao}, Chenchen and {Li}, Xinming and {Zhang}, Junshuo},
        title = "{A persistently active fast radio burst source embedded in an expanding supernova remnant}",
      journal = {Science Bulletin},
         year = 2026,
        month = jan,
       volume = {71},
       number = {1},
        pages = {76-82},
          doi = {10.1016/j.scib.2025.11.023},
       adsurl = {https://ui.adsabs.harvard.edu/abs/2026SciBu..71...76N}
}

@ARTICLE{Spitler2018,
       author = {{Spitler}, L.~G. and {Herrmann}, W. and {Bower}, G.~C. and {Chatterjee}, S. and {Cordes}, J.~M. and {Hessels}, J.~W.~T. and {Kramer}, M. and {Michilli}, D. and {Scholz}, P. and {Seymour}, A. and {Siemion}, A.~P.~V.},
        title = "{Detection of Bursts from FRB 121102 with the Effelsberg 100 m Radio Telescope at 5 GHz and the Role of Scintillation}",
      journal = {\apj},
         year = 2018,
        month = aug,
       volume = {863},
       number = {2},
          eid = {150},
        pages = {150},
          doi = {10.3847/1538-4357/aad332},
archivePrefix = {arXiv},
       eprint = {1807.03722},
 primaryClass = {astro-ph.HE},
       adsurl = {https://ui.adsabs.harvard.edu/abs/2018ApJ...863..150S}
}

@ARTICLE{Takahashi2021,
       author = {{Takahashi}, Ryuichi and {Ioka}, Kunihito and {Mori}, Asuka and {Funahashi}, Koki},
        title = "{Statistical modelling of the cosmological dispersion measure}",
      journal = {\mnras},
         year = 2021,
        month = apr,
       volume = {502},
       number = {2},
        pages = {2615-2629},
          doi = {10.1093/mnras/stab170},
archivePrefix = {arXiv},
       eprint = {2010.01560},
 primaryClass = {astro-ph.CO},
       adsurl = {https://ui.adsabs.harvard.edu/abs/2021MNRAS.502.2615T}
}

\clearpage

\end{document}